\documentclass[11pt,twocolumn,tight,times]{aastex62}
\usepackage{graphicx,color}
\usepackage{mathrsfs,amsmath}
\usepackage{ulem}
\emergencystretch=\maxdimen
\def\ergs{\rm erg~s^{-1}}

\def\feii{Fe {\sc ii}}

\def\hb{H$\beta$}
\def\hr{H$\gamma$}

\def\kms{\rm km~s^{-1}}

\def\msun{M_{\odot}}

\def\tauHb{\tau_{_{\rm H\beta}}}

\newcommand{\broad}{\ifmmode {\xi} \else $\xi$ \fi}

\def\calA{{\cal A}}
\def\calW{{\cal W}}
\def\calWW{{{\cal W}_C}}

\defcitealias{du2014}{Paper~I} 
\defcitealias{wang2014a}{Paper~II}
\defcitealias{hu2015}{Paper~III} 
\defcitealias{du2015}{Paper~IV} 
\defcitealias{du2016a}{Paper~V}   
\defcitealias{du2016b}{Paper~VI} 

\begin{document}

\title{SUPERMASSIVE BLACK HOLES WITH HIGH ACCRETION RATES IN ACTIVE GALACTIC NUCLEI. VII. \\
	RECONSTRUCTION OF VELOCITY-DELAY MAPS BY MAXIMUM ENTROPY METHOD}

\correspondingauthor{Pu Du, Jin-Ming Bai}
\email{dupu@ihep.ac.cn, baijinming@ynao.ac.cn}

\author{Ming Xiao}
\affiliation{Yunnan Observatory, Chinese Academy of Sciences, Kunming 650011, Yunnan, China}
\affiliation{Key Laboratory for Particle Astrophysics, Institute of High Energy Physics,
	Chinese Academy of Sciences, 19B Yuquan Road, Beijing 100049, China}

\author{Pu Du}
\affiliation{Key Laboratory for Particle Astrophysics, Institute of High Energy Physics,
	Chinese Academy of Sciences, 19B Yuquan Road, Beijing 100049, China}

\author{Keith Horne}
\affiliation{SUPA Physics and Astronomy, University of St.~Andrews, KY16 9SS, Scotland, UK}

\author{Chen Hu}
\affiliation{Key Laboratory for Particle Astrophysics, Institute of High Energy Physics,
	Chinese Academy of Sciences, 19B Yuquan Road, Beijing 100049, China}

\author{Yan-Rong Li}
\affiliation{Key Laboratory for Particle Astrophysics, Institute of High Energy Physics,
	Chinese Academy of Sciences, 19B Yuquan Road, Beijing 100049, China}

\author{Ying-Ke Huang}
\affiliation{Key Laboratory for Particle Astrophysics, Institute of High Energy Physics,
	Chinese Academy of Sciences, 19B Yuquan Road, Beijing 100049, China}

\author{Kai-Xing Lu}
\affiliation{Yunnan Observatory, Chinese Academy of Sciences, Kunming 650011, Yunnan, China}

\author{Jie Qiu}
\affiliation{Key Laboratory for Particle Astrophysics, Institute of High Energy Physics,
	Chinese Academy of Sciences, 19B Yuquan Road, Beijing 100049, China}

\author{Fang Wang}
\affiliation{Yunnan Observatory, Chinese Academy of Sciences, Kunming 650011, Yunnan, China}

\author{Jin-Ming Bai}
\affiliation{Yunnan Observatory, Chinese Academy of Sciences, Kunming 650011, Yunnan, China}

\author{Wei-Hao Bian}
\affiliation{Physics Department, Nanjing Normal University, Nanjing 210097, China}

\author{Luis C. Ho}
\affiliation{Kavli Institute for Astronomy and Astrophysics, Peking University, Beijing 100871, China} 
\affiliation{Department of Astronomy, School of Physics, Peking University, Beijing 100871, China} 

\author{Ye-Fei Yuan}
\affiliation{Department of Astronomy, University of Science and Technology of China, Hefei 
	230026, China}

\author{Jian-Min Wang}
\affiliation{Key Laboratory for Particle Astrophysics, Institute of High Energy Physics,
	Chinese Academy of Sciences, 19B Yuquan Road, Beijing 100049, China}
\affiliation{University of Chinese Academy of Sciences, 19A Yuquan Road, Beijing 100049, China}
\affiliation{National Astronomical Observatories of China, Chinese Academy of Sciences,
	20A Datun Road, Beijing 100020, China}

\collaboration{(SEAMBH collaboration)}

\received{2017 September 17}
\revised{2018 May 8}
\accepted{2018 July 23}
\journalinfo{To appear in {\it The Astrophysical Journal}.}

\begin{abstract}

As one of the series of papers reporting on a large reverberation mapping campaign, we 
apply the maximum entropy method (MEM) to 9 narrow-line Seyfert 1 galaxies with super-Eddington 
accretion rates observed during 2012-2013 for the velocity-delay maps of their H$\beta$ and 
H$\gamma$ emission lines. The maps of 6 objects are reliably reconstructed using MEM.
The maps of H$\beta$ and H$\gamma$ emission lines of Mrk 335 indicate that the gas of its broad-line 
region (BLR) is infalling. For Mrk 142, its H$\beta$ and H$\gamma$ lines show signatures of outflow. 
The H$\beta$ and H$\gamma$ maps of Mrk 1044 demonstrate complex kinematics -- a virialized motion 
accompanied by an outflow signature, and the H$\beta$ map of IRAS F12397+3333 is consistent with a 
disk or a spherical shell. The \hb\ maps of Mrk~486 and MCG~+06-26-012 suggest the presence of an 
inflow and outflow, respectively. These super-Eddington accretors show diverse geometry and 
kinematics. Brief discussions of their BLRs are provided for each individual object. 
\end{abstract}


\section{Introduction}
Broad emission lines are prominent features in ultraviolet and optical spectra of active galactic nuclei 
(AGNs) and are believed to stem from the so-called broad-line regions (BLRs), which are photoionized by 
the ionizing radiation from accretion disks surrounding the central supermassive black holes
\citep{osterbrock1989}. The broad emission lines reverberate in response to the varying ionizing 
continuum with a light-traveling time delay, therefore, appropriate analysis of reverberation properties 
of broad emission lines delivers information on the kinematics and geometry of BLRs (e.g., 
\citealt{bahcall1972,blandford1982,peterson1993}). Velocity-resolved time-lag analysis measures the time 
lags as a function of line-of-sight velocity. In practice, it divides the emission line into several 
velocity bins and carries out cross-correlation analysis of the light curves at different velocities. 
This is a preliminary step offering a glimpse into the geometry and kinematics of the BLRs, and has been 
applied to a number of objects
(e.g., \citealt{bentz2008,bentz2009,bentz2010,denney2009a, denney2009b,denney2010,grier2013}). 
It has been shown that  BLRs have diverse geometry and kinematics  
(such as outflows, inflows or virialized motion, see \citealt{gaskell1988,grier2013}; \citealt{du2016b}, 
hereafter \citetalias{du2016b}). The signature of inflow/outflow is identified as the mean lag
being smaller on the red/blue wing of the line profile,	while a disk-like BLR has a symmetric pattern with
smaller lags on both the red and blue wings. However, the velocity-resolved time-lag analysis measures the 
mean time lags at different velocities rather than revealing the detailed response features of broad 
emission lines. A velocity-delay map resolves the response of broad emission line not only at different 
velocities but also at different time-delay, therefore embodying all the information on the BLR response 
to the varying continuum \citep{bentz2010,pancoast2012,grier2013,pancoast2014b,grier2017}. The maximum 
entropy method (MEM, \citealt{horne1991,horne1994}, see details in Section \ref{sec3}) and dynamical 
modeling method (\citealt{pancoast2011,pancoast2014a,li2013}) were developed for this purpose. 

\begin{figure*}
	\centering
	\includegraphics[angle=0,width=0.7\textwidth]{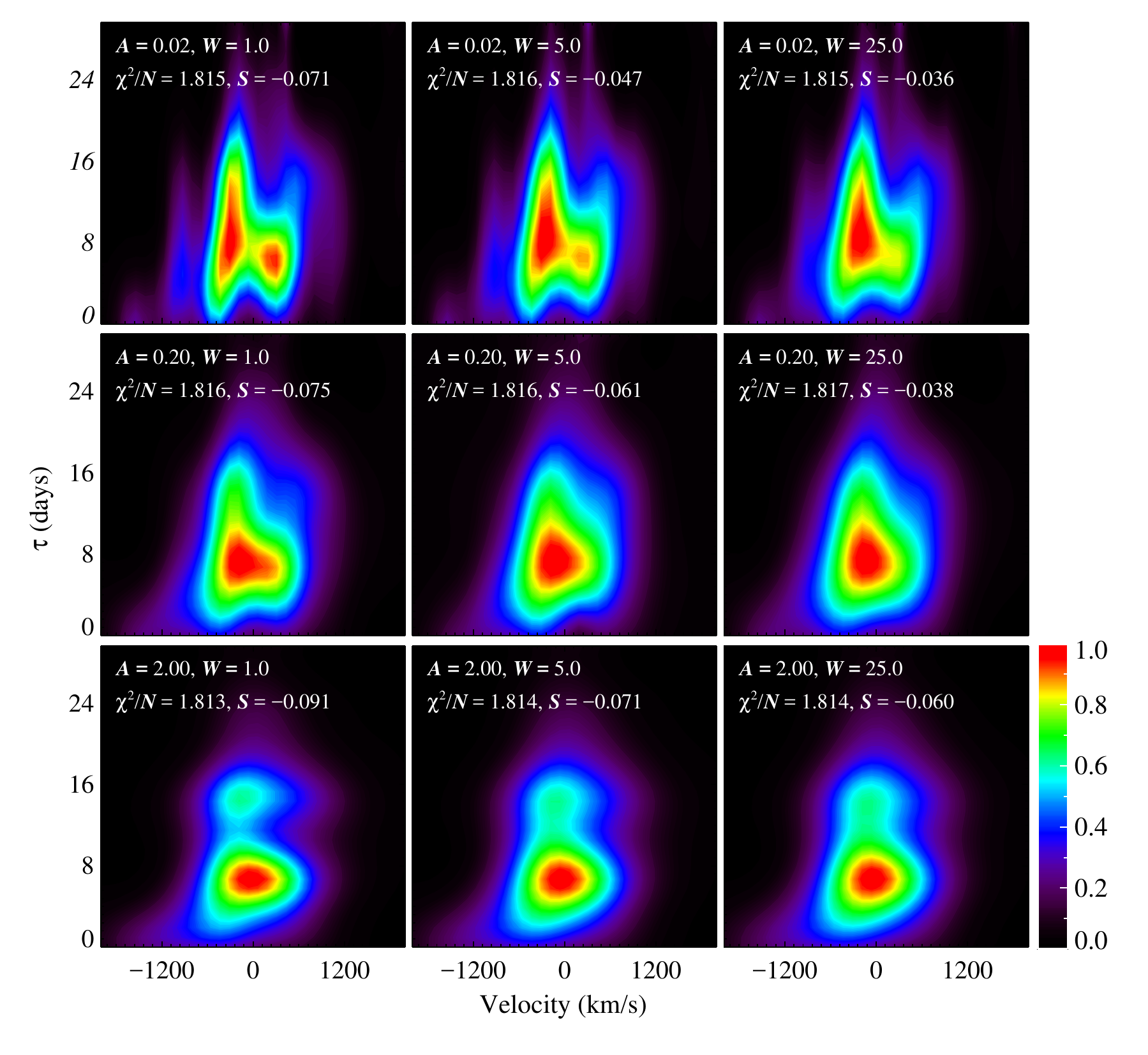} 
	\caption{\footnotesize
		H$\beta$ velocity-delay maps of Mrk~142 for different values of $\calA$~and $\calW$ ($\alpha$ 
		is fixed to 2000). From the left to right panels, the value of $\calW$ increases by factors of 
		5 while the delay maps show less flexibility.
		From the top to bottom panels, the increasing $\calA$ by factors of 10 progressively strengthens 
		the weight of entropy in the velocity direction, and smears the sub-structure in this direction. 
		The values of $\calA$~and $\calW$ in the central panel are our choice.
		The color bar is shown at the lower-right corner.
	}
	\label{AW}
\end{figure*}

Since the fall of 2012, we have monitored a sample of high accretion rate AGNs, aiming at better 
understanding the role of accretion rates on BLRs (\citealt{du2014}, hereafter \citetalias{du2014}; 
\citealt{du2015}, hereafter \citetalias{du2015}; \citealt{du2016a}, hereafter \citetalias{du2016a}) 
and the physics of accretion onto black holes (\citealt{wang2013}; \citealt{wang2014b}). The observations 
in the first three years show that super-Eddington accreting massive black holes (SEAMBHs), with 
$L_{\rm Bol}/L_{\rm Edd}\sim$ a few, have the following characteristics: 1) their H$\beta$ lags are 
significantly shorter than anticipated from the well-known BLR radius-luminosity relationship (e.g., 
\citealt{kaspi2000,bentz2013}), and the amount of shortening depends on the accretion 
rates (\citetalias{du2015,du2016a}), 2) their luminosities turn  out to be saturated, in accordance 
with predictions of the slim disk model (\citetalias{du2015,du2016a}), 3) the \feii\ emission shows 
clear reverberation in response to the varying continuum with similar lags to those of H$\beta$ lines 
\citep[hereafter \citetalias{hu2015}]{barth2013,hu2015}. Obviously, it is necessary to investigate the 
details of the BLR kinematics in those extreme objects to see how they may differ from lower accretion 
rate AGNs.

The velocity-delay maps of the BLRs for about 11 objects have been reconstructed
in the past twenty years \citep{ulrich1996,bentz2010,pancoast2012,grier2013,pancoast2014b,grier2017}. 
However, they are mainly AGNs with ``normal'' accretion rates, $L_{\rm Bol}/L_{\rm Edd}\sim0.1$. Using the 
velocity-resolved time-lag analysis, we have probed the geometry and kinematics among the BLRs in 9 SEAMBH
candidates monitored between 2012-2013 (hereafter SEAMBH2012) in \citetalias{du2016b}, identifying both 
disk-like and inflow/outflow signatures. In this paper, we use MEM to reconstruct velocity-delay maps for 
these SEAMBHs. With the high cadence and homogeneous sampling of the observations, 
we successfully recover velocity-delay maps of the H$\beta$ line for 6 objects and of the H$\gamma$ line 
for 3 objects.

The paper is organized as follows. Section 2 briefly describes the reverberation mapping (RM) observations 
and data reduction. Section 3 presents the methodology of MEM and the details of application to RM data. 
Section 4 summarizes the results of the obtained velocity-delay maps and presents discussions on individual 
objects. The conclusion is given in Section 5. Unless stated elsewise, time lags are given in the rest frame.

\section{Observations and Data}
\label{sec:ob}
In the first year of the campaign, 10 narrow-line Seyfert~1 galaxies (e.g., 
\citealt{osterbrock1985}) were selected as SEAMBH candidates and monitored from October 2012 to June 2013 
(see Table 1 in \citealt{wang2014a}, 
hereafter \citetalias{wang2014a}). H$\beta$ time lags were detected for 9 objects.
The details of telescope, spectrograph, observation and data reduction 
can be found in \citetalias{du2014}. Here we provide a brief description for completeness. 

The spectroscopic monitoring
was carried out using the 2.4~m telescope with the Yunnan Faint Object Spectrograph and Camera (YFOSC)
at the Yunnan Observatories, from the Chinese Academy of Sciences. The spectral flux was calibrated by 
observing a nearby comparison star simultaneously with the target through orienting the long
slit (2\arcsec.5 wide and 10\arcmin~long). The SEAMBH2012 observations had the following properties:
1) the redshifts of the targets range from 0.017 to 0.089;
2) the sampling is relatively high (interval average $\lesssim2$ days);
3) the signal-to-noise ratios (S/N) are high (S/N$\sim$50-100);
4) the H$\beta$ lags ($\tauHb$ $\lesssim 20$ days) are generally much shorter than 
the $\sim80$ to $\sim150$ day monitoring period;
5) the 3800-7200\AA\ wavelength coverage allows us to recover the velocity-delay maps for 
different broad emission lines (e.g, H$\gamma$ and H$\beta$); 
6) the spectra of the comparison stars 
are used to determine the line-broadening caused by instrument and seeing in each individual epoch; 
7) the {\feii} emission and the host galaxy contamination have been subtracted by a fitting scheme 
(\citetalias{hu2015}).

One point merits a further emphasis. 
The line-broadening function describes the broadening introduced by the instrument (which should not 
change much from epoch to epoch) and the variable seeing (which is different for each individual exposure), 
and widens the emission line profiles, especially for objects with relatively narrow emission lines 
(e.g., SEAMBHs). Crucially,	because a low-resolution grism and wide slit were used during our observations 
(see more details in \citetalias{du2016b}), the line-broadening of the SEAMBH2012 observations 
is, on average, $\sim500\ \kms$ (ranges from $\sim350\ \kms$ to $\sim800\ \kms$; see Figure 3 in 
\citetalias{du2016b}). The detailed line-broadening functions for different exposures can be obtained by 
de-convolving the spectra of the comparison stars in the slit. In \citetalias{du2016b},
the Richardson-Lucy deconvolution method was adopted to compensate for the line-broadening 
effect. In our MEM fitting, we convolve the model with the line-broadening function $\xi (\lambda)$ 
obtained in \citetalias{du2016b}: 
\begin{equation}
L_{\rm b}(\lambda,t_k)=L(\lambda,t_k)\otimes\xi (\lambda)
\label{broaden}
\end{equation} 
to fit the observed spectra. Here $L(\lambda,t_k)$ is the MEM modeling emission-line profile defined in 
the next section and $L_{\rm b}(\lambda,t_k)$ is the broadened MEM model.

\begin{figure*}
	\centering
	\includegraphics[angle=0,width=1\textwidth]{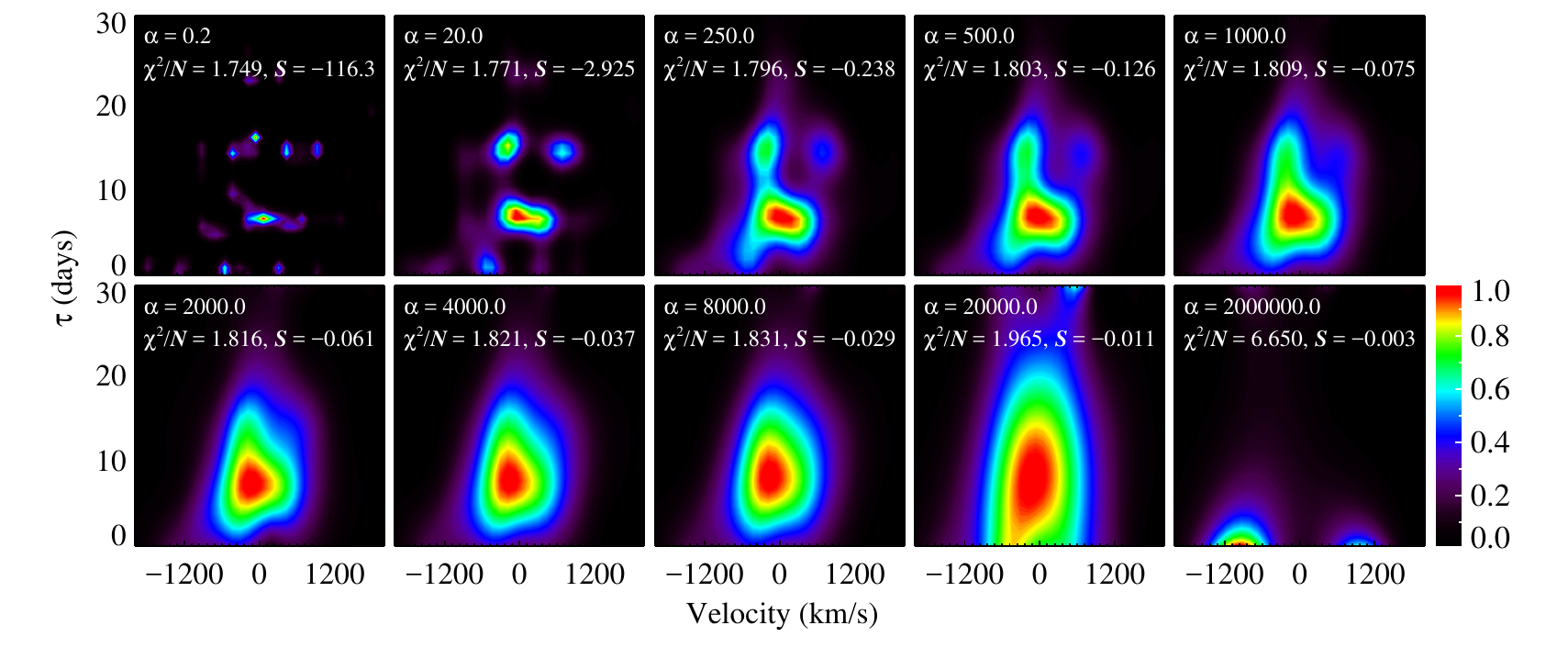} 
	\caption{\footnotesize
		H$\beta$ velocity-delay maps of Mrk~142 for different values of $\alpha$ ($\calA$ and $\calW$ are 
		fixed to 0.2 and 5, respectively). We normalize the maps and make their peak values equal to 1. 
		This normalization is also adopted for the velocity-delay maps hereafter.
		From the upper-left to the lower-right panels, we increase the value of $\alpha$. 
		The velocity-delay map becomes smoother and $\chi^2/ N$ increases as $\alpha$ increases. 
		The left bottom panel ($\alpha = 2000$) is our best choice for Mrk 142 because the 
		corresponding map is reasonably smooth and the $\chi^2/ N$ value is acceptable.
		The color bar is shown on the rightmost side.}
	\label{alpha}
\end{figure*}

\section{The maximum entropy method}
\label{sec3}

The MEM is a powerful technique 
for data fitting analysis without considering any specific model. It 
balances the $\chi^2$, which quantifies the goodness of fitting,
and the entropy $S$, which represents the smoothness and simplicity of the model.
The application of MEM to RM data was first developed by \cite{horne1991}.
Details such as the principle and equations are discussed in \cite{horne1994}. 
We generally follow the procedures of \cite{horne1994}, but with some minor changes. 
For the sake of completeness, we describe the MEM and our new  
modifications below. 

Basically, the MEM seeks a solution by minimizing
\begin{equation}
Q={\chi}^{2}-\alpha S,
\label{eq:entropy6}
\end{equation}
where the Lagrange multiplier $\alpha\geqslant0$ is introduced to trade off between ${\chi}^{2}$ and $S$. 
An undervalued $\alpha$ leads to a noisy fitting result (specifically the velocity-delay map in the following 
sections) and over-fitting of the data, whereas an overvalued $\alpha$ leads to an over-smoothed result and an 
under-fitting of the data.

In MEM, the light curves of continuum and emission line, denoted as $C(t)$ 
and $L(\lambda,t)$, respectively, are related by the formula
\begin{equation}
L(\lambda,t)=\bar{L}(\lambda)+\int_{\tau_{\rm min}}^{\tau_{\rm max}}\varPsi(\lambda,\tau)\left[C(t-\tau)-\bar{C}\right]d\tau, 
\label{eq:tf1}
\end{equation}
where $\bar{C}$ and $\bar{L}(\lambda)$ are the background levels of the light curves of the continuum and 
emission line, respectively. They account for the non-variable components in the light curves (see the details 
in \citealt{horne1994}). $\varPsi(\lambda,\tau)$ is the velocity-delay map, which is a function of time lag 
$\tau$ and line-of-sight velocity or wavelength (i.e. $\lambda$) (e.g. \citealt{bentz2010}, 
\citealt{pancoast2012}, \citealt{grier2013}). For a line at rest wavelength $\lambda_0$, the corresponding 
line-of-sight velocity is $v=c\left(\lambda-\lambda_0\right)/\lambda_0$. 
To apply MEM in RM data, we discretize equation (\ref{eq:tf1}):
\begin{equation}
L(\lambda_{i},t_{k})=\bar{L}(\lambda_i)+\sum_{j}\varPsi(\lambda_{i},\tau_{j})\left[C(t_{k}-\tau_{j})-\bar{C}\right]\Delta \tau.
\label{eq:tf3}
\end{equation}
Here, $\bar{L}(\lambda_i)$, $C(t_{k}-\tau_{j})$, and $\varPsi(\lambda_{i},\tau_{j})$ are to be determined.
The primary principle underlying the MEM is to reconstruct the ``simplest" $\varPsi(\lambda,\tau)$ 
(\citealt{skilling1984,horne1991,horne1994}) at a condition of acceptable fitting to the observed spectra.

\begin{deluxetable}{lccccccc}
	\tablecolumns{8}
	\setlength{\tabcolsep}{3pt}
	\tablewidth{0pc}
	\tablecaption{MEM Parameters\label{tab:alphatable}}
	\tabletypesize{\footnotesize}
	\tablehead{
		\colhead{Object}                &
		\colhead{$\alpha$}                &
		\colhead{$\calA$}       &
		\colhead{$\calW$}       &
		\colhead{$\tau_{\rm min}$}       &
		\colhead{$\tau_{\rm max}$}       &
		\colhead{$\chi^2/ N$}       &
		\colhead{$N$}      \\ \cline{5-6}
		& & & & \multicolumn{2}{c}{(days)} 
	}
	\startdata
	Mrk~335 (\hb)		&800		&0.5		&1	&	-10     &38        &1.489	&6160\\
	Mrk~335 (\hr)		&1200		&0.5		&1	&	-10     &38        &1.365	&4224\\
	Mrk~1044 (\hb)		&600		&0.1		&1	&	-10     &30        &2.869	&1480\\
	Mrk~1044 (\hr)		&400		&0.1		&1	&	-10     &30        &2.034	&1554\\
	Mrk~142 (\hb)		&2000		&0.2		&5	&	-5      &33        &1.816	&6000\\
	Mrk~142 (\hr)		&250		&2.0		&5	&	-2      &33        &1.723	&4100\\
	IRAS F12397+3333 (\hb)	&160		&0.5		&1	&	-10     &22        &1.706	&1720\\
	Mrk~486 (\hb)		&500		&0.1		&1	&	-5       &63       &1.865	&2340\\
	MCG~+06-26-012 (\hb)		&1800		&1		&1	&	-5     &58         &1.518	&1632\\
	\enddata
\end{deluxetable}

\begin{figure}
	\centering
	\includegraphics[angle=0,width=0.47\textwidth]{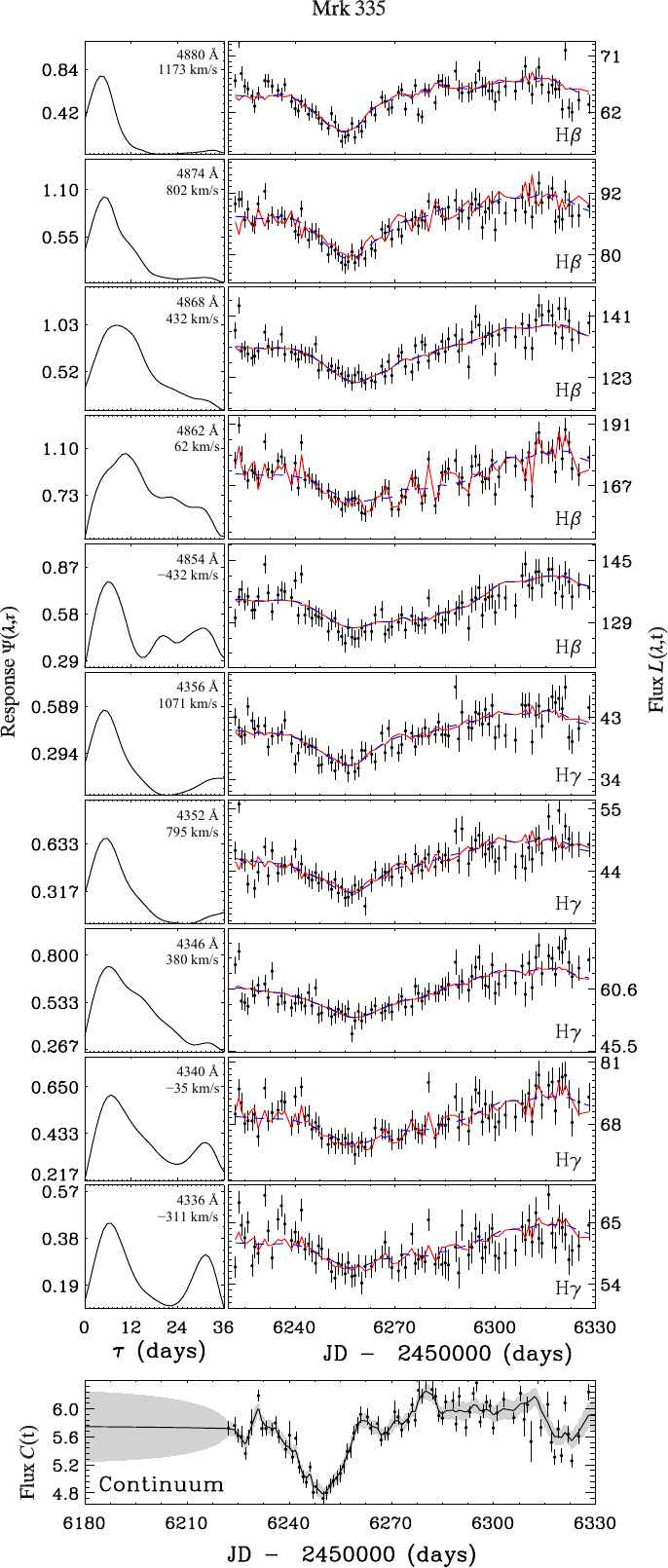} 
	\caption{\footnotesize
		The best fit to the light curves at different wavelengths (in the rest frame) for Mrk~335. 
		In each panel, the blue dashed line corresponds to the original MEM modeling light curve,
		and the solid red line denotes the light curve broadened by the line broadening function
		in each individual epoch (see $L(\lambda,t)$ and $L_{\rm b}(\lambda,t)$ in Section \ref{sec:ob}).
		The left panels show the corresponding 1-dimensional responses. The bottom panel shows the continuum 
		light curve with the reconstruction of the DRW model. The selected wavelengths are labeled along the 
		top axis of Figure \ref{VDM335}. 
		$C(t)$ and $L(\lambda, t)$ are in units of 
		$10^{-15}\ {\rm erg\ s^{-1}\ cm^{-2}\ \AA^{-1}}$ and $10^{-16}\ {\rm erg\ s^{-1}\ cm^{-2}\ \AA^{-1}}$, respectively. The response 
		$\varPsi(\lambda, t)$ is in arbitrary unit.
	}
	\label{spec335}
\end{figure}

\begin{figure}
	\centering
	\includegraphics[angle=0,width=0.47\textwidth]{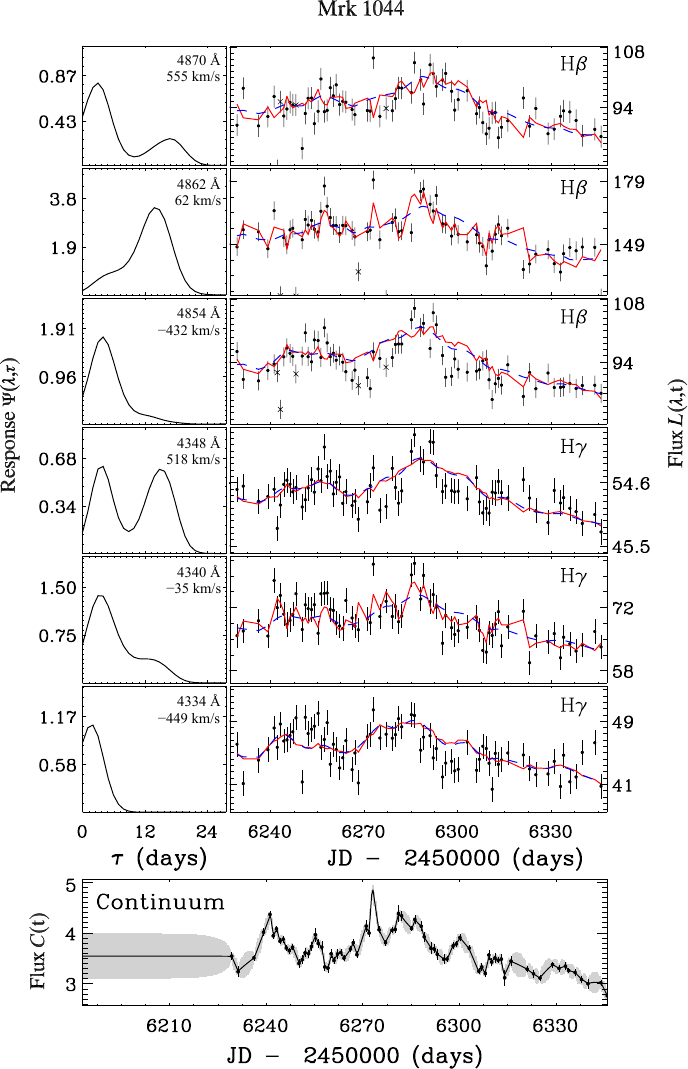} 
	\caption{\footnotesize
		Same as Figure 2 but for Mrk~1044. The selected wavelengths are labeled along the top axis of Figure 
		\ref{VDM1044}. The discarded points (see Section \ref{sec:mrk1044}) are marked with black crosses.
	}
	\label{spec1044}
\end{figure}

\begin{figure}
	\centering
	\includegraphics[angle=0,width=0.47\textwidth]{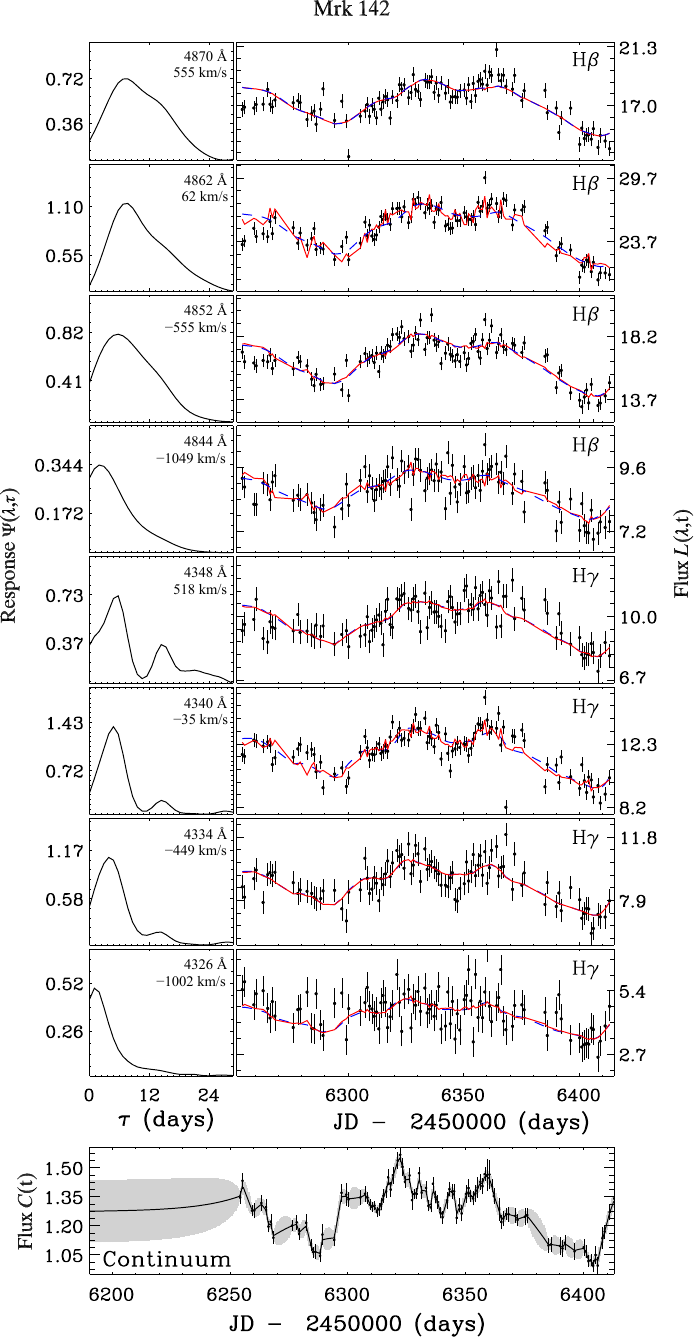} 
	\caption{\footnotesize
		Same as Figure 2 but for Mrk~142. Panels are arranged as in Figure \ref{spec335}. The selected 
		wavelengths are labeled along the top axis of Figure \ref{VDM142}.
	}
	\label{spec142}
\end{figure}

\begin{figure}
	\centering
	\includegraphics[angle=0,width=0.47\textwidth]{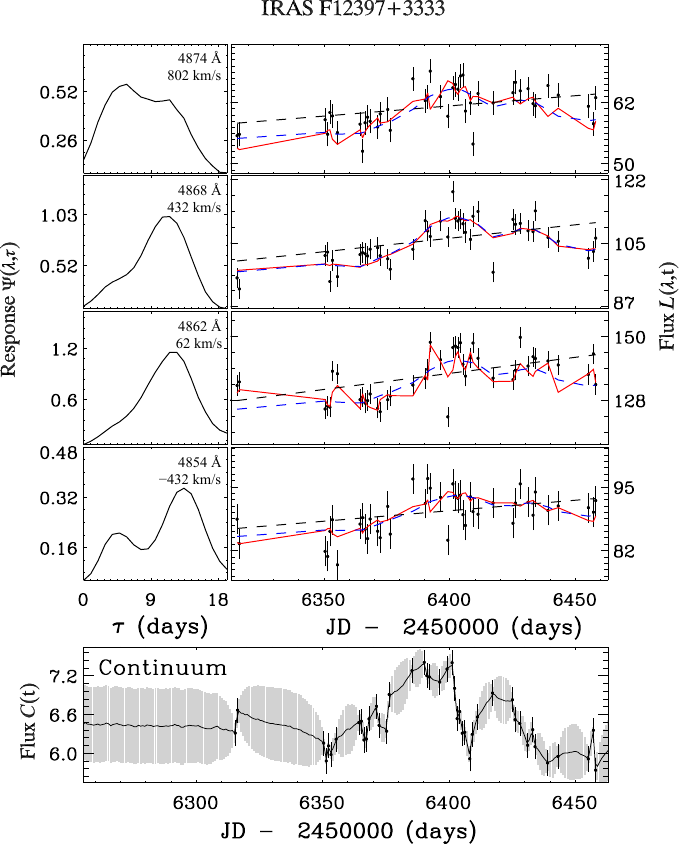} 
	\caption{\footnotesize
		Same as Figure 2 but for for IRAS F12397+3333. The black dashed lines represent the trending components 
		in the MEM model. The selected wavelengths are labeled along the top axis of Figure \ref{vdm12397}.
	}
	\label{spec12397}
\end{figure}

\begin{figure}
	\centering
	\includegraphics[angle=0,width=0.47\textwidth]{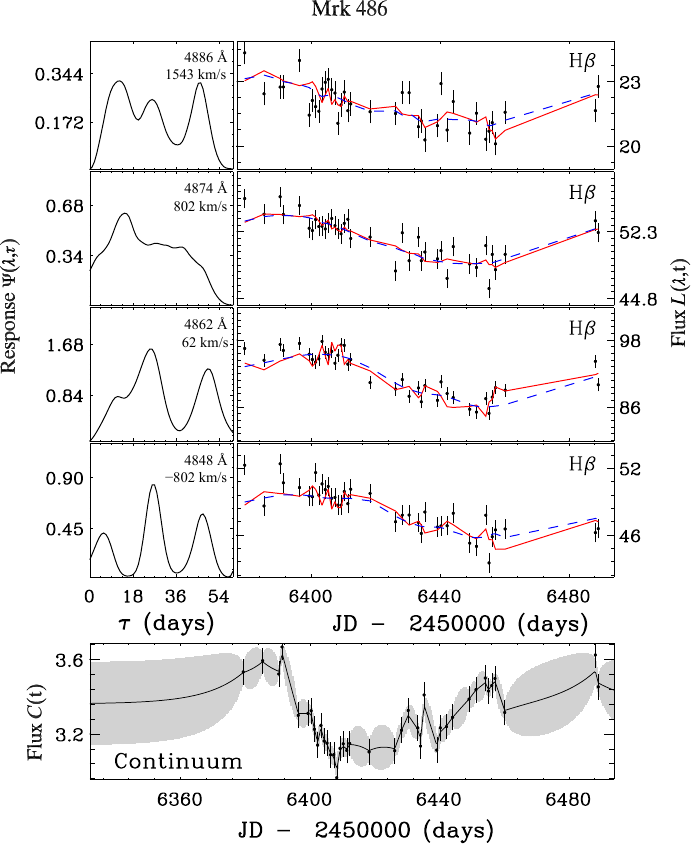} 
	\caption{\footnotesize
		Same as Figure 2 but for for Mrk~486. Panels are arranged as in Figure \ref{spec335}. The selected 
		wavelengths are labeled along the top axis of Figure \ref{VDM486}.
	}
	\label{spec486}
\end{figure}

\begin{figure}
	\centering
	\includegraphics[angle=0,width=0.47\textwidth]{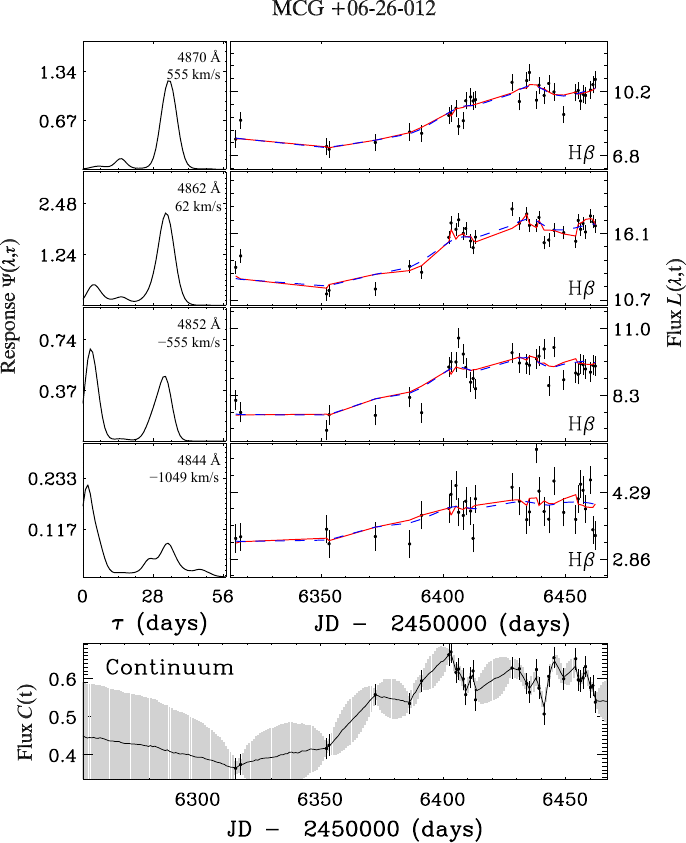} 
	\caption{\footnotesize
		Same as Figure 2 but for for MCG~+06-26-012. The selected wavelengths are labeled along the top axis 
		of Figure \ref{vdm06}.
	}
	\label{spec06}
\end{figure}

\subsection{${\chi}^{2}$: the goodness of fitting}

As in Equation (\ref{eq:tf3}), a velocity-delay map $\varPsi(\lambda,\tau)$ 
	is convolved with the continuum light curve $C(t)$ to fit observations $L_{\rm obs}(\lambda_i, t_k)$.
Here $i=1,...,N_\lambda$ and $k=1,...,N_t$ are the indexes of wavelength and epochs, respectively. 
The goodness of the fitting is evaluated by the usual $\chi^2$ as
\begin{equation}
{\chi}^{2}=\sum_{i}\sum_{k}\left[\frac{L_{\rm obs}(\lambda_{i},t_{k})- L_{\rm b}(\lambda_{i},t_{k})}{{\sigma(\lambda_{i},t_{k})}}\right]^{2}, 
\label{eq:chisquare1}
\end{equation}
where $L_{\rm b}(\lambda_{i},t_{k})$ is the broadened MEM model defined in Equation (\ref{broaden}) and 
$\sigma(\lambda_{i},t_{k})$ is the measurement error. By minimizing ${\chi}^{2}$, we obtain the closest 
fit to the data. However, the resulting $\varPsi(\lambda,\tau)$ is always very noisy, as previously described.

\subsection{$S$: the simplicity of model}
We use the definition of entropy (\citealt{horne1994}):

\begin{equation}
S=\sum_{m=1}^{M}\left[p_{m}-q_{m}-p_{m}\ln(p_{m}/q_{m})\right],
\label{eq:entropy0}
\end{equation}
where $\{{p_{m}}\}$ are the values of parameters at $M$ grid points (e.g. of the velocity-delay map
	$\Psi(\lambda,\tau)$, or of the background spectrum	$\bar{L}(\lambda)$) and 
$\{{q_{m}}\}$ are the ``default values'' which define the ``simplest" possible model. The 
$\{{q_{m}}\}$ can be regarded 
as a function of $\{{p_{m}}\}$. The entropy is valid only for $p_{m}>0$, thus it imposes 
a positivity constraint on the 
parameters. In the MEM model, $\varPsi(\lambda,\tau)$ and $\bar{L}(\lambda)$
are described by the parameters $p_{m}$.
The ``default values'' $q_{m}$ are set as a slightly blurred version of $p_{m}$ 
(\citealt{horne1994}). Following \citeauthor{horne1994} (1994), we define 
\begin{equation}
\label{eq:df1}
q(\lambda)=\sqrt{p(\lambda-\Delta\lambda)p(\lambda+\Delta\lambda)},
\end{equation}
for the background spectrum $\bar{L}(\lambda)$. 
Thus for each parameter $p_m$ the default value $q_m$
is the geometric mean of its neighbors. For the velocity-delay map $\varPsi(\lambda,\tau)$,
we define 
\begin{eqnarray}
\label{eq:df2}
\ln{q(\lambda,\tau)} &= &\dfrac{1}{ 1+\calA }\left[\ln{ \sqrt{ p(\lambda-\Delta\lambda,\tau) \, p(\lambda+\Delta\lambda,\tau) } }\right.\nonumber\\
&+&\left. \calA \, \ln{ \sqrt{ p(\lambda,\tau-\Delta\tau) \, p(\lambda,\tau+\Delta\tau) } }\right]
, 
\label{q2d}
\end{eqnarray}
where $\calA$ is a parameter that controls the relative weight of the entropies in $\tau$ and $\lambda$ 
directions. The choice for the value of $\calA$ is discussed in Section (\ref{section:aw}).	
Minimizing the $Q$ in Equation \ref{eq:entropy6} while ignoring the data is equivalent to the condition
\begin{eqnarray}
\frac{\partial S}{\partial p} = -\ln (p/q), 
\label{eq:entropy3}
\end{eqnarray}
so that maximizing $S$ gives $S=0$ at $p=q$. 
MEM draws the model toward the ``simplest'' model as defined by the default values $q$.

\begin{figure*}
	\centering
	\includegraphics[angle=0,width=\textwidth]{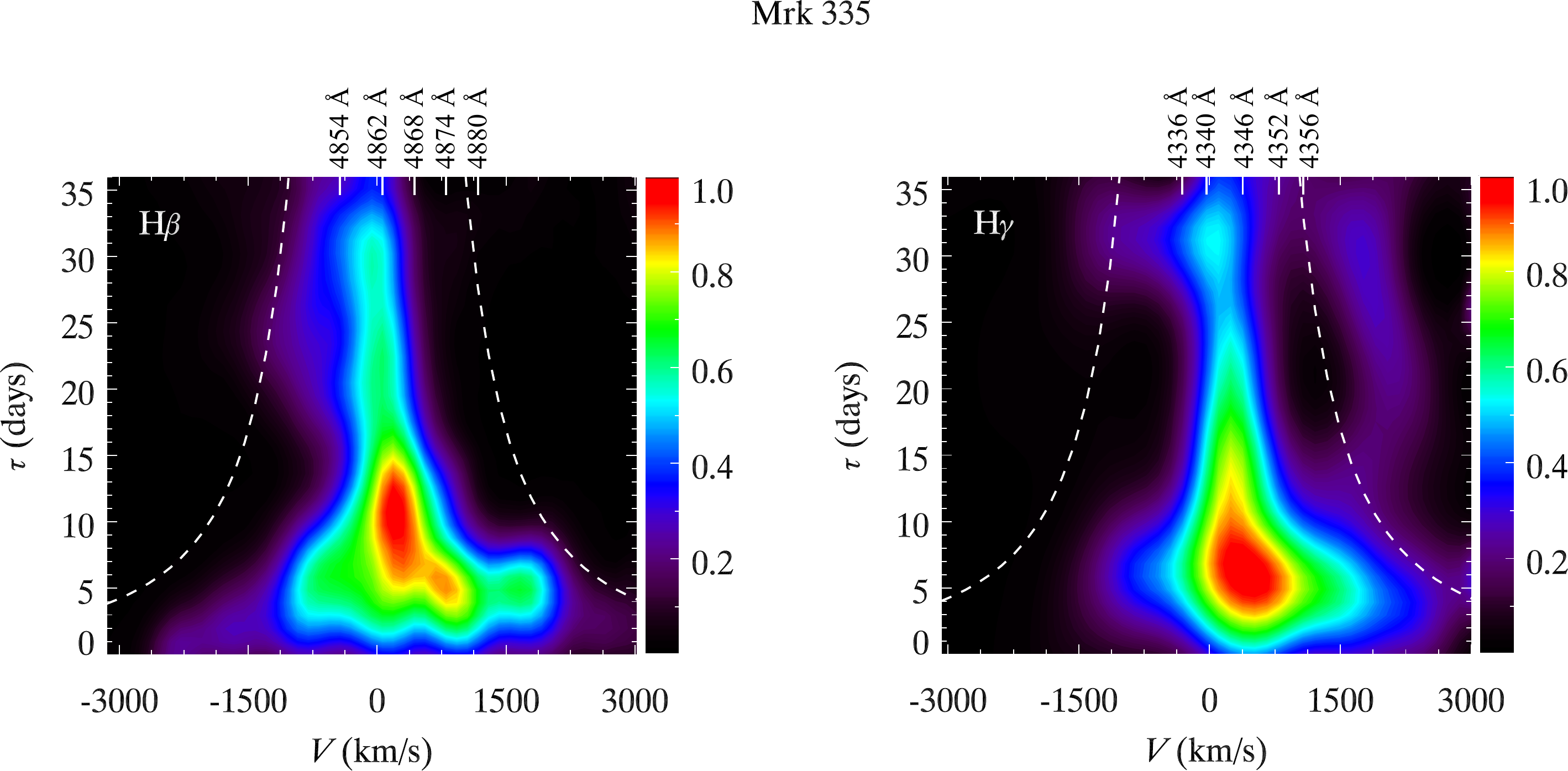}
	\caption{\footnotesize
		Velocity-delay maps of H$\beta$ and H$\gamma$~lines for Mrk~335. 
		The dashed lines show the ``virial envelope" $V^2\tau c/G = 10^{6.87}~\msun$ and the labels 
		on top axes are the corresponding rest-frame wavelengths selected in Figure \ref{spec335}. 
	}
	\label{VDM335}
\end{figure*}

\subsection{Application to RM data}
Detrending of light curves is used to obtain better detection of 
H$\beta$ lags through cross-correlation analysis, since the long-term variations possibly caused by 
the changes of BLR geometry and kinematics can affect the reverberation analysis \citep{welsh1999, denney2010, li2013}. 
By including a detrending term, Equation (\ref{eq:tf1}) is rewritten as  
\begin{eqnarray}
L(\lambda_{i},t_{k})&=&\bar{L}_{\rm trend}(\lambda_{i},t_{k})\nonumber\\
  &+&\sum_{j}\varPsi(\lambda_{i},\tau_{j})\left[C(t_{k}-\tau_{j})-\bar{C}\right]\Delta \tau, 
\label{eq:2d1}
\end{eqnarray}
where $\bar{L}_{\rm trend}(\lambda_{i},t_{k})$ describes long-term trends in the line light curves. 
We set $\bar{C}$ as the median of continuum flux.
We use a linear polynomial to represent the trending components of $\bar{L}_{\rm 
trend}(\lambda_{i},t_{k})$, namely, $\bar{L}_{\rm trend}(\lambda_{i},t_{k}) = \bar{L}(\lambda_{i})+\left[b
(\lambda_{i})-b_0\right] x(t_{k})$,
where $x(t_{k}) = 2(t_{k}-\bar{t})/(t_{\rm max}-t_{\rm min})$, $b(\lambda_{i})-b_0$ 
is the linear slope to be determined and $\bar{L}(\lambda_{i})$ is the constant 
background. Here $\bar{t}$, $t_{\rm max}$, and $t_{\rm min}$ are the mean, maximum, 
and minimum times of the light curves, respectively, and $b_0$ is a constant.
We adopt a form for $x(t_{k})$ so that it runs from -1 to +1 across the time span 
of the observations. In addition, We take 
$b(\lambda_{i})-b_0$ as the slope, since $b(\lambda_{i})$ is included in the
entropy terms and is limited to a positive number (see Equation \ref{eq:entropy0}).
We evaluate the slopes of the \hb\ line light curves of our sample, the results 
range between [-6.0 $\sim$ 8.2] $\times10^{-16}\ergs\rm cm^{-2} \AA^{-1}$, thus we take $b_0 
	= 10$ ($\times10^{-16}\ergs\rm cm^{-2} \AA^{-1}$), in which
the slopes of the trending components $b(\lambda_{i})-b_0$
can range from $-10$ to $+\infty$ ($\times10^{-16}\ergs\rm cm^{-2} \AA^{-1}$), 
and the result of our fitting can be fully optimized.
We find that, except for IRAS 
F12397+3333, the slopes of the trending components in the other five objects are approximately zero. 
We thus exclude the detrending term in the fitting of these five objects to reduce 
the number of unknown parameters, and only take into account the detrending term in IRAS F12397+3333. 
For IRAS F12397+3333, the linear slope $\left[b(\lambda_{i})-b_0\right]/(t_{\rm max}-t_{\rm min})$ ranges 
between [0.007 $\sim$ 0.063]$\times10^{-16}\ergs\rm cm^{-2} \AA^{-1}\rm days^{-1}$, and the average is 
0.029$\times 10^{-16}\ergs\rm cm^{-2} \AA^{-1}\rm days^{-1}$ (see the fitting of its light curves shown in 
Section \ref{sec:vdm}, and $t_{\rm max}-t_{\rm min} = 142$ days for IRAS F12397+3333).

The total entropy of the model can be written as
\begin{eqnarray}
S=\frac{S_{\bar{L}}+S_{b}+\calW S_{\varPsi}}{1+\calW},
\label{eq:2d2}
\end{eqnarray}
where $S_{\bar{L}}$, $S_{b}$, and $S_{\varPsi}$ are the entropies
of the background, slope, and velocity-delay map, respectively.
Each term has the form shown in Equation \ref{eq:entropy0}. The corresponding ``default value''
$q$ in $S_{\bar{L}}$ and $S_{b}$ follows Equation \ref{eq:df1}, and that of $S_{\varPsi}$ follows 
Equation \ref{eq:df2}. The parameter $\calW$ controls the weight of the entropy from the last term 
relative to the first two terms in the numerator of the right-hand side of Equation (\ref{eq:2d2}). 
We take $S_{\bar{L}}$ and $S_{b}$ as equal weight terms, which is different to $S_{\varPsi}$,
as $\bar{L}$ and $b$ are both parts of the background spectrum and are in the same units 
(as described above). We changed the weight of these two terms from 0.1 to 10, to test that the 
general structure of the final velocity-delay maps is robust.
How to select the best values for parameter $\calW$ and the $\calA$ (defined in Equation \ref{q2d}) 
and their influences on the velocity-delay maps are demonstrated in \cite{horne1994} 
and \cite{grier2013}. We follow their procedures and show the resulting velocity-delay maps for 
several potential sets of $\calA$, $\calW$ in Figure \ref{AW}.

The continuum term $C(t_k-\tau_j)$ in Equation (\ref{eq:2d1}) requires interpolation among the 
discrete measured data points. Instead of following the procedure in \cite{horne1994}, in which 
the continuum light curve is treated as a component of the model, here we first interpolate the 
continuum light curve $C(t)$ at any given time point using a damped random walk (DRW) model 
(e.g., \citealt{li2013, zu2013}) and do not let it vary throughout the MEM fitting, only 
adjusting the velocity-delay map and the background spectra (and their trends).
Thus the entropy related to the continuum light curves does not appear in 
Equation (\ref{eq:2d2}). It should be noted that fixing $C(t)$ indeed does not fully optimize 
$C(t)$, since the information in the $L(\lambda, t)$ light curves does not feed back into the 
determination of $C(t)$. However, our results should not be strongly affected because the 
sampling of SEAMBH2012 is high (averagely less than $\sim$2 days). In order to evaluate the 
influences of different continuum modeling, we compare the output maps for two approaches: 
one approach is adopting a simple linear interpolation of the observed continuum light curve 
(Appendix \ref{appa}) and the other is jointly modeling continuum and emission line by 
including the entropy of continuum  into the total entropy (Appendix \ref{appb}).
These results show that the different continuum modeling do not remarkably influence our resulting
velocity-delay maps. Employing the DRW model greatly reduces the number of parameters in our model.
\subsection{Control parameters $\alpha$, $\calA$ and $\calW$}
\label{section:aw} 
The parameter $\alpha$ 
influences the final resolution of the map and plays a role in the quality of the MEM fits. Figure 
\ref{alpha} illustrates this effect. With $\alpha$ increasing, $\chi^2/ N$ increases but the 
resolution becomes lower (i.e., the fine-scale structures in the velocity-delay map disappear). 
However, if the value of $\alpha$ is too low, the $\chi^2$ term dominates the entropy term in Equation 
(\ref{eq:entropy6}). As a result, the more flexible model allows MEM to overfit noise in the light 
curves. For each individual objects, we choose the best $\alpha$ by checking the fitting of light 
curves at different wavelengths and the final resolution of the map. This is similar to the 
procedure of \cite{grier2013}.

\begin{figure*}
	\centering
	\includegraphics[angle=0,width=\textwidth]{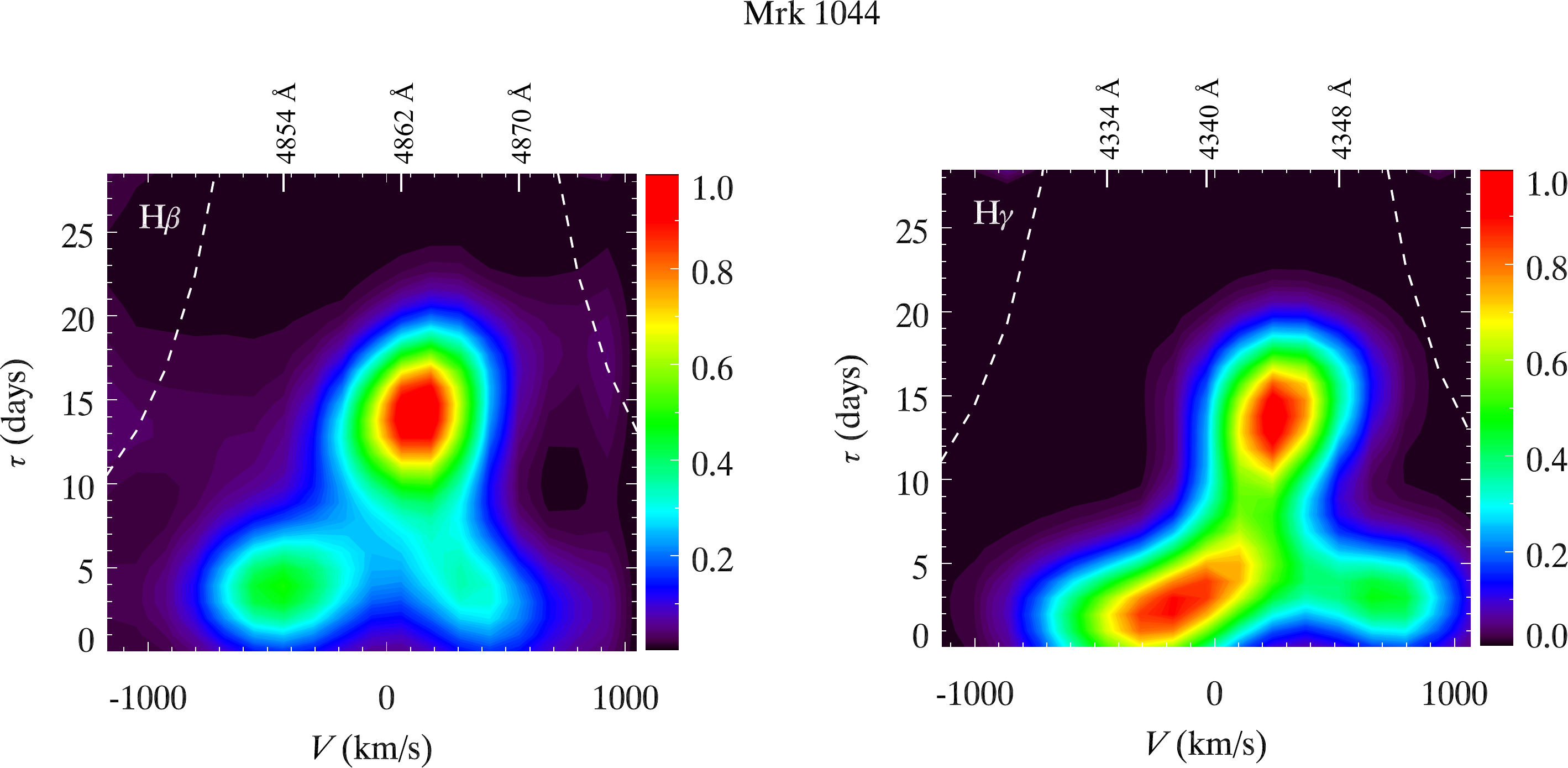}
	\caption{\footnotesize
		Velocity-delay maps of H$\beta$ and H$\gamma$~lines for Mrk~1044. 
		The dashed lines show the ``virial envelope" $V^2\tau c/G = 10^{6.45}~\msun$ and the labels 
		on top axes are the corresponding rest-frame wavelengths selected in Figure \ref{spec1044}.}
	\label{VDM1044}
\end{figure*}

As mentioned above, $\calW$ controls the weight of the entropy of the velocity-delay map relative 
to those of $b(\lambda)$ and $\bar{L}(\lambda)$. $\calA$ controls the ratio of the entropy in
velocity and $\tau$ direction. Due to the differences in the emission line widths and the 
observation cadences, the best values of $\calA$~and $\calW$ are different object-to-object. 
For the purpose of illustration, we plot the resulting velocity-delay maps of Mrk~142 in Figure 
\ref{AW}, in which we alter the value of $\calA$~and $\calW$. Increasing $\calW$ (by factors of 
5 from the left to right panels) makes the map smoother and smoother. Increasing the value of 
$\calA$ (from top to bottom panels) smears out fine structures along the velocity direction, 
because the weight of entropy in this direction increases. In general, the fits with different 
$\calA$ and $\calW$ are similar in the sense that the $\chi^2/ N$ does not change significantly 
(much less than the influence of $\alpha$). This means that the MEM procedures are robust and 
relatively insensitive to $\calA$~and $\calW$. Here we adopt $\calA$ = 0.2 and $\calW$ = 5 for 
the H$\beta$ map of Mrk~142 as the best one, because too small or too large ($\calA$, $\calW$) 
may introduce spurious small structure or make the map over-smoothed. 

\section{The velocity-delay maps}
\label{sec:vdm}
\begin{figure*}
	\centering
	\includegraphics[angle=0,width=\textwidth]{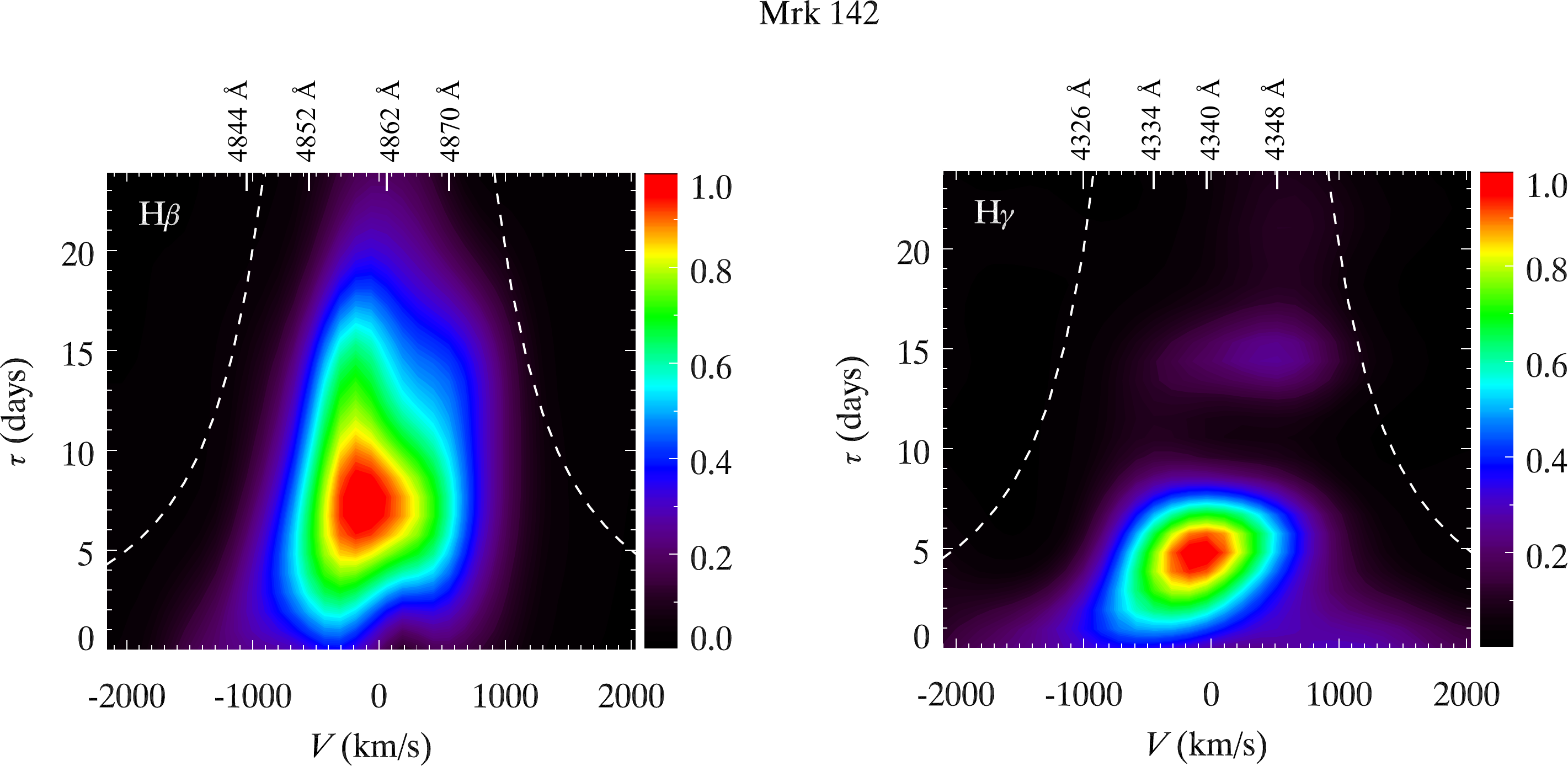}
	\caption{\footnotesize
		Velocity-delay maps of H$\beta$ and H$\gamma$~lines of Mrk~142. 
		The dashed lines show the ``virial envelope" $V^2\tau c/G = 10^{6.59}~\msun$ and the labels 
		on top axes are the corresponding rest-frame wavelengths selected in Figure \ref{spec142}.}
	\label{VDM142}
\end{figure*}

We apply MEM to the 9 SEAMBH candidates with significant lag measurements (8 of them are identified 
as SEAMBHs except for MCG~+06-26-012). The fitting to the light curves of emission lines at some 
selected wavelengths and the corresponding one-dimensional response functions are presented in 
Figures \ref{spec335}-\ref{spec06}. The MEM modeling light curves are relatively ``noisy'' after 
convolving with the line-broadening function, since the function varies each night. 
Although the individual spectra are smoother after the convolution, the light curves are noisier.
For a better illustration, we plot both the original MEM modeling light curves \footnote{For a 
better comparison, we convolve the original MEM modeling spectra with a Gaussian kernel which has 
the mean FWHM of the line-broadening functions at all epochs.} 
and the broadened ones in Figure \ref{spec335}-\ref{spec06}.
The DRW reconstructions of the continuum light curves are also plotted in Figures 
\ref{spec335}-\ref{spec06}. As can be seen, the variation features of light curves at different 
wavelengths are generally well reproduced, indicating that our MEM procedure is finding 
models that	adequately fit light curve variations in the data of the present SEAMBH sample.
The resulting velocity-delay maps are shown in Figures \ref{VDM335}-\ref{vdm06}, 
and their corresponding $\alpha$, $\calA$, $\calW$, $\chi^2/N$, $N$ and the lower/upper limit 
$\tau_{\rm min}$/$\tau_{\rm max}$ of the delay in 
the MEM calculation are listed in Table \ref{tab:alphatable}. The selected wavelengths
in Figures \ref{spec335}-\ref{spec06} are also marked along the top axis. 
In general, the velocity-delay maps of the 5
objects (Mrk~335, Mrk~142, Mkr~1044, Mrk~486, and MCG~+06-26-012) are in good agreement with 
the velocity-resolved time lags. However, due to inclusion of the detrending term,
the velocity-delay map of IRAS F12397+3333 shows slightly shorter time lags than the 
velocity-resolved time-lags from cross-correlation analysis.
MCG~+06-26-012 was selected as a SEAMBH candidate in our sample, but subsequent work revealed 
that it is a sub-Eddington object (\citetalias{wang2014a,du2015}).

In Figures \ref{VDM335}-\ref{vdm06}, we also plot the ``virial envelope" $V^2=GM_\bullet/c\tau$ for 
each object, where $G$ is the gravitational constant, $M_\bullet$ is black hole mass, and $c$ is the 
speed of light. We adopt the values of black hole mass derived in \citetalias{du2015}, as listed also 
in the figure captions. In general, all of the observed features in the velocity-delay maps for the 
six objects are found within their corresponding envelopes. In principle, the response of a virialized 
BLR is symmetric and wider (narrower) at shorter (longer) delays and should be confined within the 
envelope. In contrast, the BLRs of outflow and inflow show asymmetric signatures in their 
velocity-delay maps. Simulated velocity-delay maps were first presented in \cite{welsh1991} and more 
recently in Figure~10 of \cite{bentz2009} and Figure~14 of \cite{grier2013}. These provide good 
references for the structures found from the observations. Below we present discussions on the results 
for individual objects. Several outliers (one point in a spectrum of Mrk~142\footnote{The outlier of 
Mrk~142 deviates very far away from the other points in the light curves, and is not visible in Figure 
\ref{spec142}.}, and several points in five spectra for Mrk~1044 shown in Figure \ref{spec1044} that 
may influence the velocity-delay maps were traced	to bad CCD pixels and then removed from the analysis.

\subsection{Mrk~335.}
\label{section:mrk335} 
We apply MEM to the data of Mrk~335 obtained in SEAMBH2012 and show its H$\beta$ and H$\gamma$ 
velocity-delay maps in Figure \ref{VDM335}, with light curve fits at selected wavelengths in Figure 
\ref{spec335}. Both the H$\beta$ and H$\gamma$ maps exhibit response over a wider range of $\tau$ 
(0 to 30 days) in the line core than in the line wings (0 to 10 days). Equivalently, the response at 
small $\tau$ extends over a wider velocity range than at larger $\tau$. The red wing response is 
stronger than that	in the blue wing, and the ``red-leads-blue'' asymmetric pattern is the expected 
signature of inflow.

Mrk~335 had been monitored spectroscopically several times in the past three decades (the campaign of 
1989 - 1996 in \citealt{kassebaum1997}, the campaign of 2010 - 2011 in \citealt{grier2012}, and SEAMBH2012 
in \citetalias{du2014}). Its velocity-delay map in the campaign of 2010 - 2011 \citep{grier2012} was 
provided using the two different methods (MEM in \citealt{grier2013} and dynamical modeling in 
\citealt{grier2017}), and shows very significant signature of inflow, which is manifested as longer 
lags in the blue wing and shorter lags towards the red wing. A similar inflow feature was evident in 
the campaign of SEAMBH2012 using velocity-resolved analysis (\citetalias{du2016b}).

Our result is generally consistent with the kinematic 
signature of inflow. However, compared with the map (mainly at $\sim15$~-~$35$ days) in the campaign of 
2010 - 2011 obtained by \cite{grier2013}, the response of H$\beta$ in 2012 - 2013 becomes faster 
($\sim2$~-~$25$ days) and shows a potential trend of extending 
towards the corner of blueshift velocities and shorter lags (a trend from [$200~\kms$, 10 days] to 
[$-1300~\kms$, 2~days] at $\sim$50\% of the maximum response level of the map). 
This may imply that the BLR of Mrk~335 is evolving from inflow to an inclined disk or a spherical shell. 
This is also supported by the lag changing from $14.3^{+0.7}_{-0.7}$ days (\citealt{grier2013}) to
$8.7^{+1.6}_{-1.9}$ days (\citetalias{du2016b}). However, the 5100~\AA\ luminosity does not change 
significantly (from $43.74\pm0.06$ to $43.69\pm0.06$ $\ergs$). 
If we assume that the ionization parameter $U=L_{\rm ion}/4\pi R^2 cn_{\rm H}h\nu$ is a constant 
(this assumption is supported by the observation of 
the BLR radius-luminosity relationship, see \citealt{koratkar1991} and \citealt{peterson1993}), 
where $L_{\rm ion}$ is the ionizing luminosity, $R$ is the radius of the BLR, $c$ is the speed of light, 
$n_{\rm H}$ is the hydrogen number density, $h$ is the Planck constant, and $\nu$ is the frequency of the 
ionizing photons. Clearly, this model shows that the changes in $R$ are not due to variations in 
$L_{\rm ion}$. Considering that two years have passed since the campaign of \cite{grier2012}, it
is not unexpected that some dynamical changes began to happen in the BLR of Mrk~335.  
The map of H$\gamma$ is almost the same as that of H$\beta$, implying that the H$\beta$ and 
H$\gamma$ regions have similar kinematics and geometry.

\subsection{Mrk~1044.}
\label{sec:mrk1044}
In Figure \ref{VDM1044}, we show the H$\beta$ and H$\gamma$ velocity-delay maps of Mrk~1044. 
The maps exhibit three connected features. The strongest one is at 
$\tau\sim15$~days and $v\sim+200$~km~s$^{-1}$. Of the two weaker ones with $\tau<5~$days and 
$v\sim\pm500$~km~s$^{-1}$, the stronger is in the blue wing, suggesting outflow.

In \citetalias{du2016b}, the velocity-resolved analysis divides the H$\beta$ emission line of 
Mrk~1044 into 4 wavelength bins and shows longer lags at smaller velocities and shorter lags at 
higher velocities. Such symmetric feature was interpreted as evidence of virialized motion.
The H$\beta$ velocity-delay map shows similar structures.
In general, the BLR of Mrk 1044 tends to be virialized. However, the overall structure of the map 
is accompanied by an outflow signature, with shorter response towards the blue end and longer response 
occurring in the red end (the average delay is 8 days for $<$ 0 $\kms$, and 11 days for $>$ 0 $\kms$). 
The response of the blue blob at $[-500~\kms, 3~days]$ is stronger than the red one at $[500~\kms, 3~days]$.

The \hr~map is generally similar to the \hb~map, but
its strongest response extends to shorter lags. It means that more \hr-emitting gas is located at smaller radius.
The \hr~map shows a more definite outflow signature, although the red blob at $[700~\kms, 3~days]$ still exists. 
This provides an indirect support to the outflow signature found in the \hb~map
(considering that the \hb~and \hr~lines have the same emitting mechanism).

As an AGN with extremely high accretion rate (its dimensionless accretion rate  
$\mathscr{\dot{M}}=\dot{M}_{\bullet}\,c^2/L_{\rm Edd}\sim16.6$ where $\dot{M}_{\bullet}$ is 
the accretion rate and $L_{\rm Edd}$ is the Eddington luminosity; 
see \citetalias{du2015}), 
the relatively stronger radiation pressure, compared with 
the normal AGNs, would be a likely driver of the outflow component of the BLR. Moreover, 
the self-shadowing effect of slim accretion disks in SEAMBHs \citep{wang2014b} can make 
the situation more complicated. The geometrically thick funnel of the inner region of a 
slim disk naturally leads to two dynamically distinct regions of the BLR, namely, shadowed and unshadowed regions 
\citep{wang2014b}. The self-shadowing effect can be a possible reason for the very complex structure found in the 
\hb~and \hr~map of Mrk~1044. Quantitative details of the self-shadowing effect on the BLR of Mrk~1044 need careful 
calculations and simulations are needed to quantitatively study the self-shadowing effects on the BLR of Mrk 1044. 
We defer this to a separated paper.

\begin{figure}
	\centering
	\includegraphics[angle=0,width=0.48\textwidth]{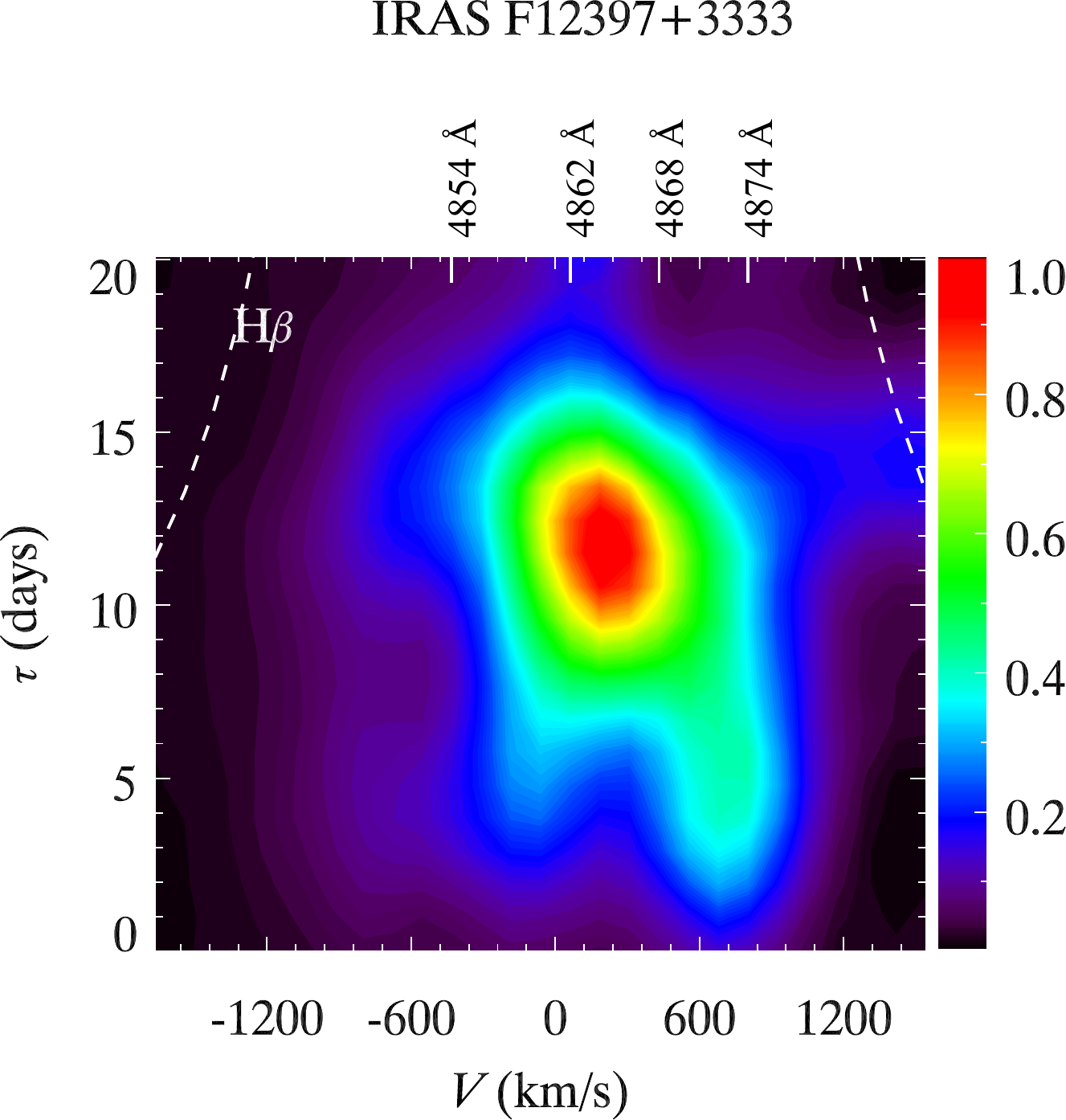} 
	\caption{\footnotesize
		Velocity-delay maps of H$\beta$ line of IRAS F12397+3333. 
		The dashed lines show the ``virial envelope" $V^2\tau c/G = 10^{6.79}~\msun$ and the labels 
		on top axis are the corresponding rest-frame wavelengths selected in Figure \ref{spec12397}.}
	\label{vdm12397}
\end{figure}

\subsection{Mrk~142.}
In the velocity-resolved time-lag analysis of \citetalias{du2016b}, the longer lags in the red end and shorter 
lags in the blue end indicate that the BLR of Mrk~142 tends to be outflowing. Our obtained velocity-delay maps 
of H$\beta$ and H$\gamma$ using MEM show more details and are illustrated in Figure \ref{VDM142}. The 
strongest response of H$\beta$ is located from $\sim5$ to $\sim11$ days at zero velocity. The 
velocity-resolved results from \citetalias{du2016b} are almost the same. 
It divides the \hb~line into 8 bins, and the time lags at the line center are $\sim9$ days.
It should be noted that the velocity-resolved time lags are the average delays of the response in the 
corresponding velocity bins of the map.
The response of H$\gamma$ peaks at a shorter $\sim5$~day lag, with a weaker feature at $\sim 15$~days.

\begin{figure}
	\centering
	\includegraphics[angle=0,width=0.48\textwidth]{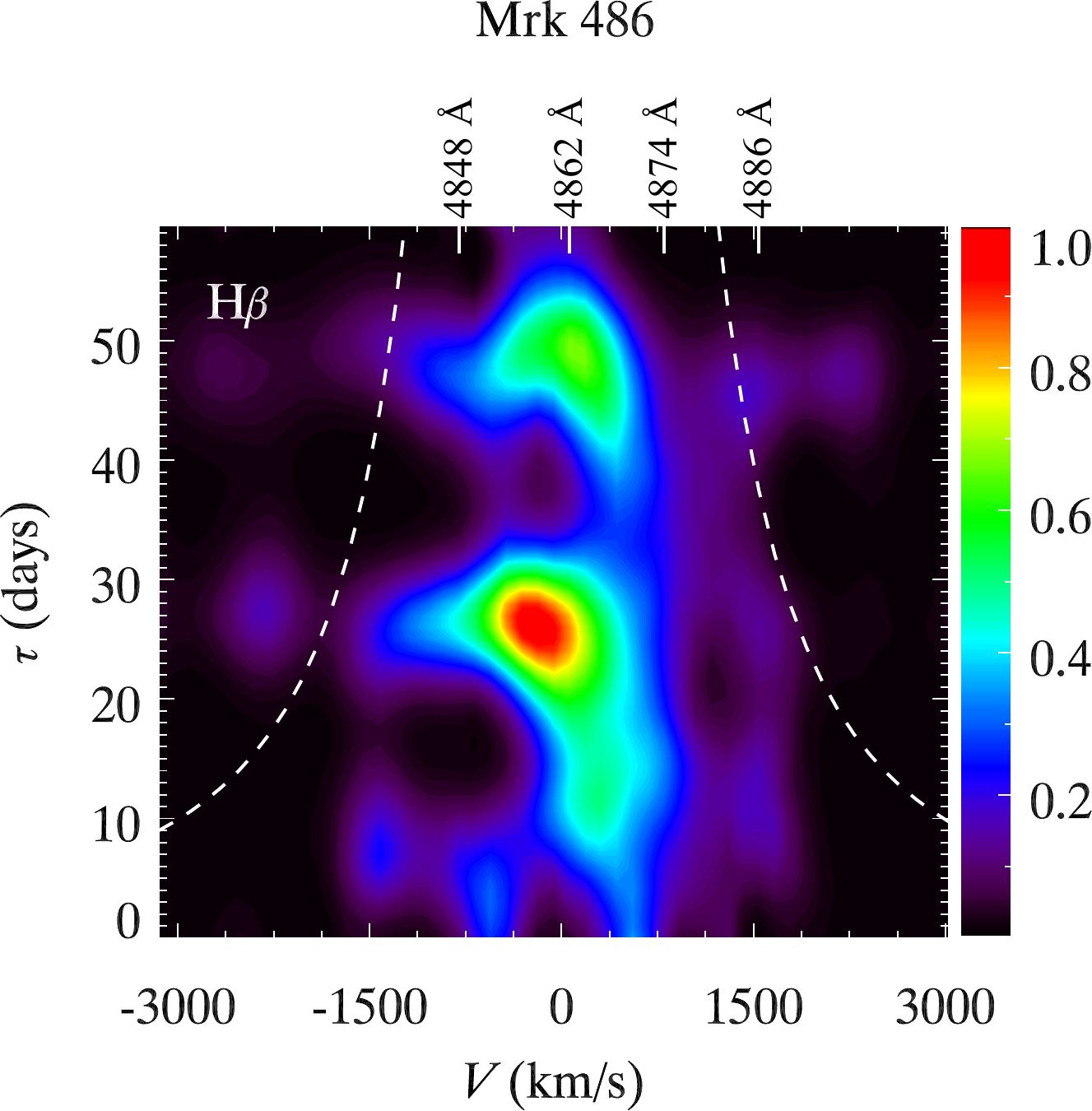} 
	\caption{\footnotesize
		Velocity-delay maps of H$\beta$ line of Mrk~486. 
		The dashed lines show the ``virial envelope" $V^2\tau c/G = 10^{7.24}~\msun$ and the labels 
		on top axis are the corresponding rest-frame wavelengths selected in Figure \ref{spec486}.}
	\label{VDM486}
\end{figure}

\subsection{IRAS F12397+3333.}
\label{sec:12397}
In the velocity-resolved time-lag analysis for IRAS F12397+3333 in \citetalias{du2016b}, 
the time lags at different velocity bins are almost indistinguishable. 
However, we take into account the long-term trend in the emission-line light curves (see Equation
(\ref{eq:2d1})) and reconstruct successfully the velocity-delay map of its H$\beta$ in Figure \ref{vdm12397}. 
The long-term trend in our emission-line model is shown in Figure \ref{spec12397} as the black dashed line 
for each wavelength. Compared with the velocity-resolved time lags
in \citetalias{du2016b} (which ranges from $\sim12$ to $\sim16$ days, except for only one bin), 
the velocity-delay map obtained here has slightly shorter time lags (the strongest response is located at 
$\sim$12 days), ascribed to the detrending of the emission line. The velocity-delay map is symmetric and 
consistent with the structure of an inclined disk or spherical shell \citep{grier2013}. 
This is similar to the case of 3C~120 in \cite{grier2013} or Mrk~50 in \cite{pancoast2012}.
\subsection{Mrk~486 and MCG~+06-26-012.}
\begin{figure}
	\centering
	\includegraphics[angle=0,width=0.48\textwidth]{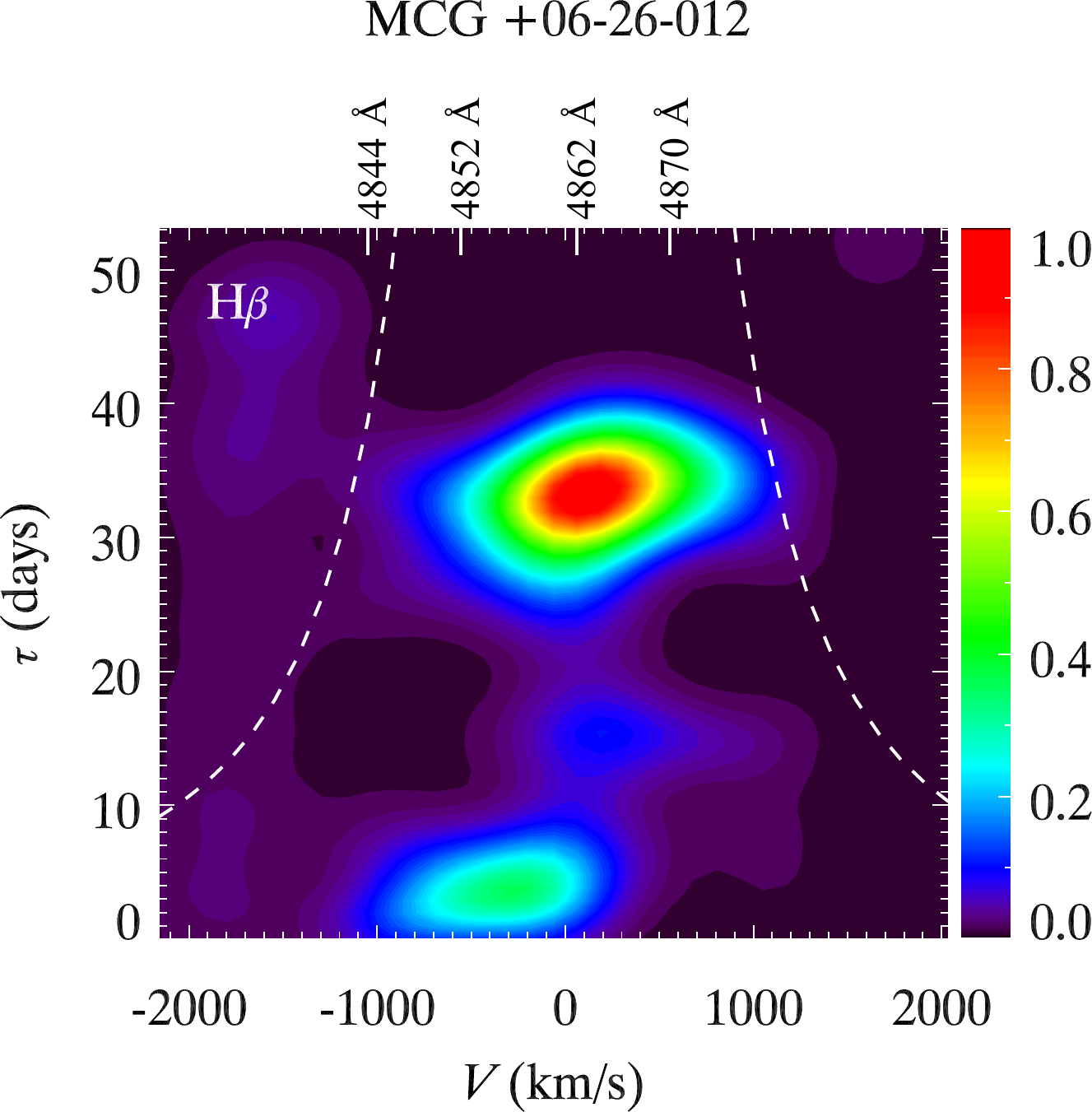} 
	\caption{\footnotesize
		Velocity-delay maps of H$\beta$ line for MCG~+06-26-012. 
		The dashed lines show the ``virial envelope" $V^2\tau c/G = 10^{6.92}~\msun$ and the labels 
		on top axis are the corresponding rest-frame wavelengths selected in Figure \ref{spec06}.}
	\label{vdm06}
\end{figure}

The velocity-delay maps of Mrk~486 and MCG~+06-26-012
show signatures of inflow (Figure \ref{VDM486}) and outflow (Figure \ref{vdm06}), respectively, consistent 
with the velocity-resolved lags in \citetalias{du2016b}. The response from 35 to 55 days in the map of Mrk~486 
is uncertain to some extent, because of the large gap from Julian day (hereafter JD) 2,456,460 to 2,456,488 in 
the \hb~light curves (see Figure \ref{spec486}). But it should be noted that the longest 
response can not exceed $\sim60$ days
(the last two points in each \hb\ light curves have already 
raised up, which means that the dip is not later than $\sim$ JD 2,456,488, while
the dip of the continuum is at $\sim$ JD 2,456,430).
The weak feature from 0 to 7 days in the map of MCG~+06-26-012 is less reliable, due to the gaps from 
JD 2,456,315 to 2,456,386 in the \hb~light curves (see Figure \ref{spec06}).

\subsection{Mrk~382, Mrk~493, and IRAS 04416+1215.}
We fail to reconstruct the velocity-delay maps of H$\beta$ emission lines for Mrk~382, Mrk~493, 
and IRAS 04416+1215, because of the relatively smaller variation amplitudes. 
Their maps should be obtained with future observations.
\section{SUMMARY}
We have reconstructed the velocity-delay maps for 6 objects in our SEAMBH2012 campaign. The BLR of Mrk 335 
is consistent with an infalling dynamics, maintaining a similar structure as in the campaign
of \cite{grier2012} two years ago. Mrk 486 also shows inflow signatures in its velocity-delay map. 
Mrk 1044 shows a complex pattern that looks like a combination of an outflow 
and a symmetric component. These two components may be generated from the self-shadowing effect 
of slim accretion disk in SEAMBH \citep{wang2014b}. 
The maps for Mrk 142 and MCG +06-26-012 also show features suggestive of an outflow, but for 
IRAS F12397+3333, the map is consistent with a disk-like geometry or a spherical shell. 
The velocity-delay maps obtained by MEM provide a basis for the model selection in dynamical modeling method 
in our follow-up works.

\acknowledgements{
We acknowledge the support of the staff of the Lijiang 2.4m telescope. Funding for 
the telescope has been provided by CAS and the People's Government of Yunnan Province. This research is 
supported by National Key R$\&$D Program of China (grant 2016YFA0400701), by Grant No. QYZDJ-SSW-SLH007 
from the Key Research Program of Frontier Sciences, CAS, and by NSFC through grants NSFC-11173023, 
-11133006, -11373024, -11233003, -11473002, and -11573026. K.H. acknowledges support from STFC grant 
ST/M001296/1 and ST/R000824/1. The work of LCH was supported by the National Key R$\&$D Program of China 
(2016YFA0400702) and the National Science Foundation of China (11473002, 11721303).

}

\appendix

\section{Interpolation for continuum light curve}
\label{appa}
\begin{figure*}
	\centering
	\includegraphics[angle=0,width=0.9\textwidth]{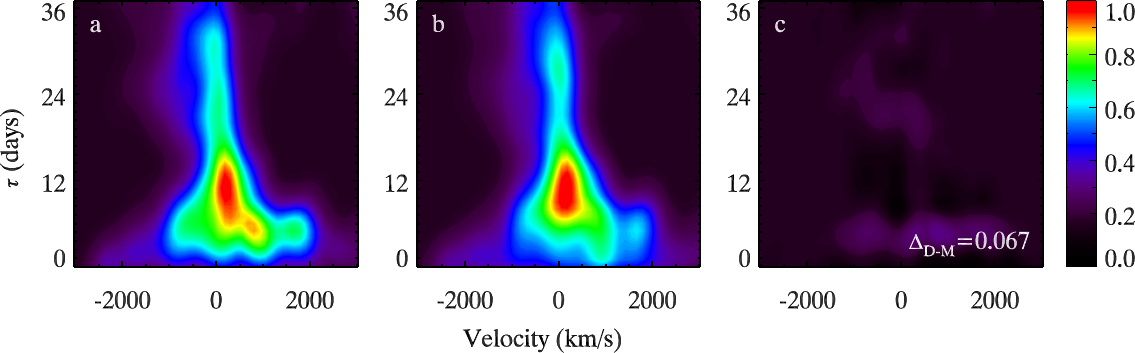} 
	\caption{\footnotesize
		Comparison between the $\varPsi_{\rm D}(\lambda_{i},\tau_{j})$ and $\varPsi_{\rm L}(\lambda_{i},\tau_{j})$ of 
		Mrk~335. Panel a shows
		$\varPsi_{\rm D}(\lambda_{i},\tau_{j})$, panel b shows $\varPsi_{\rm L}(\lambda_{i},\tau_{j})$, and panel c 
		is the residual	$(\varPsi_{\rm D}-\varPsi_{\rm L})$. 
	}
	\label{appalinterp}
\end{figure*}

\begin{figure*}
	\centering
	\includegraphics[angle=0,width=0.9\textwidth]{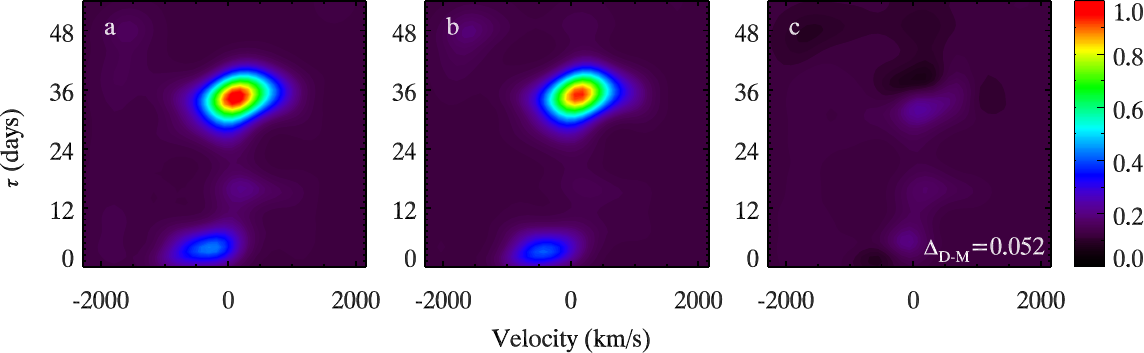} 
	\caption{\footnotesize
		Comparison between the $\varPsi_{\rm D}(\lambda_{i},\tau_{j})$ and $\varPsi_{\rm L}(\lambda_{i},\tau_{j})$ of 
		MCG~+06-26-012. Panel a shows
		$\varPsi_{\rm D}(\lambda_{i},\tau_{j})$, panel b shows $\varPsi_{\rm L}(\lambda_{i},\tau_{j})$, and panel c 
		is the residual $(\varPsi_{\rm D}-\varPsi_{\rm L})$. 
	}
	\label{appa06}
\end{figure*}

In the present paper, we interpolate the continuum light curves by using DRW.
Whether DRW is a good description of the light curves in SEAMBHs remains to be 
investigated (the DRW model may not be appropriate for all AGNs, see \citealt{kasliwal2015}), 
but this is beyond the scope of the present work. Details will be discussed in a forthcoming paper 
(Lu et al. 2018, in preparation). We investigate here how the output velocity-delay maps are affected 
if we use simple linear interpolation for the continuum light curves. We demonstrate the comparison of 
the reconstructed \hb~maps of Mrk~335 and MCG~+06-26-012 as two examples in Figure \ref{appalinterp} 
and \ref{appa06} for the cases of DRW-interpolated and linear-interpolated continuum light curves. The 
maps are denoted as $\varPsi_{\rm D}(\lambda_{i},\tau_{j})$ and $\varPsi_{\rm L}(\lambda_{i},\tau_{j})$, 
respectively. We adopt the same values of the MEM parameters listed in Table \ref{tab:alphatable} for the
reconstruction of the velocity-delay maps.  The difference between the two maps can be quantified by  
\begin{eqnarray}
\Delta_{\rm D-L}=\frac{\sqrt{\sum_{i,j}\left[\varPsi_{\rm D}(\lambda_{i},\tau_{j})-\varPsi_{\rm L}(\lambda_{i},\tau_{j})\right]^2/N}}
{\left(\bar{\varPsi}_{\rm max}+\bar{\varPsi}_{\rm min}\right)/2},
\end{eqnarray}
where $\bar{\varPsi}(\lambda_{i},\tau_{j})=0.5\times[\varPsi_{\rm D}(\lambda_{i},\tau_{j})+\varPsi_{\rm L}(\lambda_{i},\tau_{j})]$ is
the average map of $\varPsi_{\rm D}(\lambda_{i},\tau_{j})$ and $\varPsi_{\rm L}(\lambda_{i},\tau_{j})$, and 
$\bar{\varPsi}_{\rm max}$ and 
$\bar{\varPsi}_{\rm min}$ are the maximum and minimum of $\bar{\varPsi}(\lambda_{i},\tau_{j})$, respectively. 
In general, the two maps of Mrk~335 are very similar ($\Delta_{\rm D-L}$ = 0.067) 
and show the same signatures. 
The $\Delta_{\rm D-L}$ values of the other five objects (listed in Table \ref{tab:appatable1}) are also
very small (the $\Delta_{\rm D-L}$ ranges from 0.024 to 0.083). 
This test shows that the continuum interpolation methods do not influence the output velocity-delay maps of the 
present sample very significantly.

\begin{deluxetable}{lcc}
	\tablecolumns{6}
	\tablewidth{0pc}
	\tablecaption{Differences of the map between different continuum modeling\label{tab:appatable1}}
	\tabletypesize{\footnotesize}
	\tablehead{
		\colhead{Object}                &
		\colhead{$\Delta_{\rm D-L}$}                &
		\colhead{$\Delta_{\rm D-M}$}       
	}
	\startdata
	Mrk~335 (\hb)		&0.067		&0.062\\
	Mrk~335 (\hr)		&0.071		&0.077\\
	Mrk~142 (\hb)		&0.024		&0.055\\
	Mrk~142 (\hr)		&0.037		&0.098\\
	Mrk~1044 (\hb)		&0.083		&0.088\\
	Mrk~1044 (\hr)		&0.063		&0.096\\
	IRAS F12397+3333 (\hb)	&0.029		&0.060\\
	Mrk~486 (\hb)		&0.053		&0.051\\
	MCG~+06-26-012 (\hb)		&0.052		&0.040\\
	\enddata
\end{deluxetable}

\section{Simultaneous MEM continuum modeling}
\label{appb}
\begin{figure*}
	\centering
	\includegraphics[angle=0,width=1\textwidth]{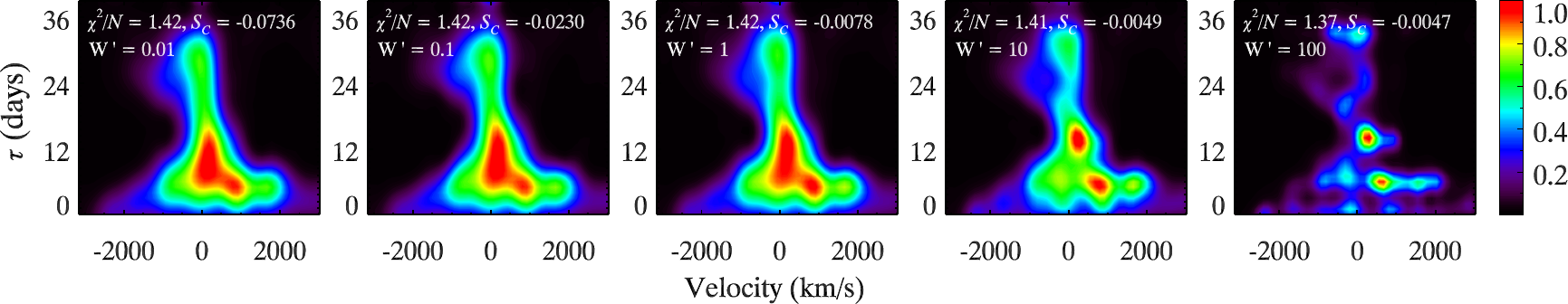} 
	\caption{\footnotesize
		The velocity-delay maps of Mrk~335 reconstructed for the cases of MEM-modeled continuum light curves. 
		From left to right, we vary $W'$ from 0.01 to 100 (see Appendix \ref{appb}). 
	}
	\label{appamrk335}
\end{figure*}

\begin{figure*}
	\centering
	\includegraphics[angle=0,width=1\textwidth]{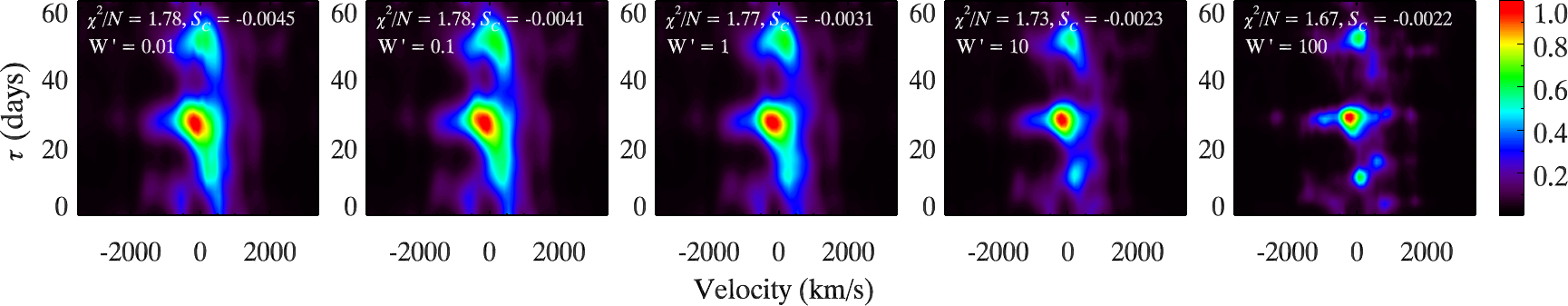} 
	\caption{\footnotesize
		The velocity-delay maps of Mrk~486 reconstructed for the cases of MEM-modeled continuum light curves. 
		From left to right, we vary $W'$ from 0.01 to 100 (see Appendix \ref{appb}). 
	}
	\label{appamrk486}
\end{figure*}

The continuum light curve can be also modeled simultaneously with the emission-line light curves using 
MEM. We investigate here how the output velocity-delay maps are affected if we adopt MEM modeling to the continuum.
We re-define Equation (\ref{eq:2d2}) as
\begin{eqnarray}
\label{2d3}
S=\frac{S_{\bar{L}}+S_{b}+\calW S_{\varPsi}+\calWW S_{C}}{1+\calW+\calWW},
\label{eq:2d3}
\end{eqnarray}
where $S_{C}$ is the entropy of the continuum and $\calWW$~is a new parameter controlling the ``stiffness'' of $C(t)$. 
Following \cite{grier2013}, we model the continuum light curves by DRW first, in order to obtain 
better constraints on the fitting, 
especially in the gaps (e.g., the gaps in the light curves of Mrk~486). We then use MEM to reconstruct the 
continuum light curves as well as the velocity-delay maps. 
As two typical examples, the 
resulting velocity-delay maps for Mrk~335 and Mrk~486 are shown in Figures \ref{appamrk335} and \ref{appamrk486}, 
respectively. We vary the $\calWW$~parameter from 0.01 to 100. 
The general features of the resulting maps are not affected by the changes of $\calWW$, and are consistent with
the results already obtained above.
The influence of the different $\calWW$ to the map of Mrk~486 is relatively stronger, because the
continuum interpolation in the large gaps of its light curve becomes more flexible if adopting MEM continuum modeling. 
Similar to Appendix \ref{appa}, we define $\Delta_{\rm D-M}$ to evaluate the difference between the velocity-delay
maps reconstructed for the cases of DRW-interpolated ($\varPsi_{\rm D}(\lambda_{i},\tau_{j})$) and MEM-modeled
($\varPsi_{\rm M}(\lambda_{i},\tau_{j})$) continuum light curves, and provide $\Delta_{\rm D-M}$ values ($\calWW$ is 
fixed as 1 for $\varPsi_{\rm M}(\lambda_{i},\tau_{j})$) for the objects in Table \ref{tab:appatable1}. 
For all of the objects, $\Delta_{\rm D-M}$ values are very small (0.040$\sim$0.098). Therefore, the inclusion of 
continuum in the MEM reconstruction does not change the velocity-delay maps significantly.

\end{document}